\documentclass[
showkeys,
preprint
amsmath,
amssymb,
amsfonts,
aps,
pre,%
groupedaddress,
superscriptaddress,
a4paper,%
nofootinbib,
eprint,
preprintnumbers, 
]{revtex4-2}

\usepackage[utf8]{inputenc}
\usepackage{fontenc}

\usepackage{subfigure}

\usepackage{multirow}
\usepackage[width=18.0cm, height=26.0cm]{geometry}
\usepackage{physics}
\usepackage{dcolumn}
\usepackage{bm}
\usepackage{graphics,graphicx} 
\usepackage{wasysym}
\usepackage{color,xcolor}
\definecolor{goldenyellow}{rgb}{0.94, 0.60, 0.01}

\usepackage[backref=page]{hyperref}

\usepackage{natbib}
\hypersetup{
	colorlinks=true,
	linkcolor=red,
	filecolor=magenta,      
	urlcolor=black,
	citecolor=cyan,
	linktocpage=true,
	breaklinks=false,
	pdftitle={Signature of chaos in perturbed quantum wells},
	pdfauthor={Pranaya Pratik Das}
}

\usepackage{booktabs}
\usepackage{siunitx}

\usepackage{soul}

\usepackage{float}

\graphicspath{{Figures/}} 

\begin{document}
\normalsize
\title{Out-of-Time-Order-Correlation in perturbed quantum wells}
	

\author{Pranaya Pratik Das}
\email{pranayapratik\_das@nitrkl.ac.in}
\author{Biplab Ganguli}%
\email{biplabg@nitrkl.ac.in}
\affiliation{%
	Department of Physics and Astronomy, National Institute of Technology Rourkela, Odisha, India-769008}%

\date{\today}

\begin{abstract}
\textit{Out-of-Time-Order-Correlator} (OTOC) and \textit{Loschmidt Echo} (LE) are commonly regarded as diagnostic tools for chaos, although they may yield misleading results because of various other factors. Previous studies have concluded that OTOC shows exponential growth in the neighbourhood of a local maximum. If this statement holds true, the exponential growth should break off once the local maximum is no longer present within the system. By applying a small symmetry-breaking perturbation, we notice that the behaviour of the OTOCs remains remarkably resilient even in the absence of a maximum. Besides this, we also notice that with the increase in perturbation strength, the broken symmetric region expands, causing a broader range of eigenstates to engage in the exponential growth of OTOCs. Therefore, the critical factor lies not in the presence of a local maximum, but in the dynamic nature of the density of states in the broken symmetry regions. Our examination, spanning diverse potential landscapes, reveals the universality of this phenomenon. We also use other chaos diagnostic tool, LE. Interestingly, it also shows signature of chaos whenever there is an exponential growth of OTOC.
\end{abstract}

\keywords{Symmetry breaking, OTOC, Loschmidt Echo, IHO,  Chaos, Quantum Chaos}

\maketitle
\pagestyle{plain}
\section{\label{sc1}Introduction}

The Lyapunov exponent, extensively discussed in notable publications\cite{WOLF1985285, S0218127418500670, F_Christiansen_1997, PhysRevLett.74.70}, effectively characterises classical chaos. It measures the divergence of nearby trajectories in phase space, which is a measure of sensitivity to initial conditions. However, the Heisenberg uncertainty principle dictates that, for a system with $N$ degrees of freedom, a single quantum state occupies a volume of $\hbar^N$ in classical phase space. Hence, we no longer have the luxury of following individual orbits. Thus, there is a pressing need for a chaos diagnostic tool tailored for quantum systems.

Several diagnostic tools are developed over the years to detect chaos in quantum systems. Some of them are level-spacing-distribution\cite{PhysRevLett.52.1, haake2019quantum}, SFF\cite{Cotler2017, PhysRevD.100.026017, PhysRevD.98.086026}, level number variance\cite{GUHR1998189, PhysRevD.98.086026, PhysRevD.100.026017}, entanglement power\cite{PhysRevE.70.016217, PhysRevA.62.030301, PhysRevE.64.036207, PhysRevA.63.040304}, quantum coherence\cite{PhysRevResearch.3.023214}, Loschmidt echo\cite{PhysRevA.30.1610, PhysRevLett.86.2490, PhysRevLett.92.150403, PhysRevLett.97.194103, doi:10.1080/00018730902831009} etc. Nonetheless, the OTOC\cite{PhysRevLett.121.210601, das2022generalised, PhysRevA.103.062214, Romatschke2021, PhysRevE.106.L062201, Lewis-Swan2019, PhysRevResearch.3.033155, PhysRevE.99.012201, PhysRevLett.118.086801, PhysRevE.107.054202, PhysRevD.101.026021, PhysRevB.96.060301, PhysRevE.100.042201} has become widely recognised and adopted in the high energy and condensed matter physics, despite initially being considered in the context of superconductivity\cite{Larkin}.

The OTOCs, within the framework of quantum mechanics, are the growth of non-commutating quantum mechanical operators describing the \textit{Unequal Time Commutation Relations} (UTCRs). It is the most potent quantum mechanical analogue of the classical sensitiveness to the initial conditions. However, the OTOC measures sensitivity in a scrambled time\footnote{Firstly, OTOC can measure a more nuanced notion of sensitivity as quantum corrections allow it to contain higher-order derivatives of $ x(t) $. Secondly, the quantum terms are dominant at the \textit{scrambling time} (\textit{Ehrenfest time}$= t_{\mathcal{E}}\sim \dfrac{1}{\lambda_L}\ln (\dfrac{k_{\beta}}{\hbar}) $), where $\lambda_L$ is the Lyapunov exponent, which explains known deviations from exponential growth at this early time scale.} order quite differently than its classical counterpart. Unlike classical systems, quantum measurements inherently disturb the measured system, and the OTOCs measure this disturbance by considering both forward and backward time evolution.

Despite its utility, the relationship between OTOC and chaos is subtle. Several researchers have explored the behaviour of OTOCs in the context of chaos and potential functions. For instance, K. Hashimoto \textit{et al.}\cite{Hashimoto2020}, R. A. Kidd \textit{et al.}\cite{PhysRevA.103.033304}, and B. Bhattacharjee \textit{et al.}\cite{bhattacharjee2022krylov} have argued that exponential growth in OTOCs near a local maximum of a potential function may not necessarily indicate chaos. Supporting this view, Kirkby \textit{et al.} \cite{PhysRevA.104.043308} also claimed that relying solely on OTOCs to detect chaos can yield false signals. Xu \textit{et al.} \cite{PhysRevLett.124.140602} further argued that exponential growth of OTOCs, often referred to as scrambling, does not inevitably imply chaos. Takeshi Morita \cite{PhysRevD.106.106001} demonstrated that OTOCs for a localised wave packet representing a classical particle near a hill in the potential function exhibit exponential growth.

The primary focus of the aforementioned works is to observe the behaviour of OTOC near a saddle point of an integrable system. Naturally, the suitable candidates for such a study is a double well potential, having an \textit{inverted harmonic potential} (IHO) term. These studies conclude that the exponential growth of the OTOC is a universal characteristic irrespective of integrability of the system, provided the potential includes a local maximum, indicating instability. This line of argument does not explore the deeper understanding of the general cause of exponential growth in OTOC. This raises an intriguing question: what happens if perturbed well does not have a local maximum? If the local maximum is the sole driver of exponential growth, then its absence should show no such behaviour in the OTOC.

To explore this, we consider one-dimensional quantum mechanical models with three specific polynomial potentials \cite{Hashimoto2017, Hashimoto2020, PhysRevD.106.106001, Garbaczewski_2022, Aquino2011EnergyEF, PhysRevE.105.034204, PhysRevA.41.1218} with a linear perturbation term. We  study the behaviour of OTOC and other chaos diagnostic tools while perturbation strength is varied. We extend the study when the saddle point disappears at a suitable perturbation. Such a perturbation introduces symmetry breaking in the system\cite{hand_finch_1998, Reichl2021, Bracken20, PhysRevA.40.438, Sun2022, PhysRevE.74.056608, DSB_Y_Nambu, PhysRevA.35.398, Song_2023}. Understanding the interplay of bifurcation and symmetry breaking is crucial in investigating phenomena like flames, chemical reactions, biological pattern formation, stability analysis, and the onset of chaos in different systems\cite{field2009symmetry, bayliss1991bifurcation, krause2024turing, CHOSSAT1988423}. 

As the saddle point disappears, the potential function becomes more asymmetric, driven by an asymmetric term (perturbation). We confirm this asymmetry in the Hamiltonian by looking at the classical bifurcation diagram and the structure of quantum eigenstates and eigenvalues\cite{Reichl2021}. Our primary objective is to quantify how asymmetry affects OTOCs: whether it amplifies, suppresses, or alters the pattern of exponential growth. To reinforce our findings, we employ additional chaos diagnostic tools, specifically the Loschmidt Echo, for all models in nearly identical scenarios. This helps validate and complement our observations of chaotic behaviour influenced by symmetry-breaking perturbations in quantum systems.

The structure of this paper is as follows. In section (\ref{sc2}), we discuss the formalism for computing the OTOC of a quantum mechanical system with a time-independent Hamiltonian. In section (\ref{sc3}), we devise three models, namely a double well, a double well with a plateau (subsection~\ref{sc3A}), and a triple well (subsection~\ref{sc3B}). In the individual subsections of the models, we perform a quantitative analysis to investigate the connection between the classical bifurcation diagram and symmetry breaking in the potential landscape, while varying different parameters. The introduction of perturbation disrupts the inherent symmetry within these potentials, leading to the emergence of asymmetric structures. Following the solution of the Schrödinger equations for these quantum mechanical models under specific perturbation strengths, we proceed to measure the microcanonical \& thermal OTOCs, and density of states. Subsection (\ref{sc3C}) contains our numerical observations on the above-mentioned three models using other chaos diagnostic tool; Loschmidt echo. Subsequently, we engage in an extended discussion of the obtained results, reinforced by additional analysis in section (\ref{sc4}). Finally, in section (\ref{sc5}), we summarise our results.

\section{\label{sc2}OTOC in quantum mechanics for a time-independent Hamiltonian}
A 2N-point OTOC \cite{das2022generalised, PhysRevResearch.3.033155} is defined as:
\begin{equation}\label{eq1}
	C_{\beta} (t_1, t_2) = -\ev{\comm{\hat{x}(t_1)}{\hat{p}(t_2)}^N} 
\end{equation}
While the $N=1$ case exhibits random but decaying behaviour, studying the $N=2$ case, involving the four-point correlator, provides a complete understanding of time disorder averaging. Exploring higher-order cases would be unnecessary for our study.

For simplicity, we take $ t_1=t $ and  $ t_2=0 $ in Eq.(\ref{eq1}), and define the four-point OTOC\cite{PhysRevE.98.062218} as: 
\begin{equation}\label{eq2}
	\mathcal{C}_{\beta}(t)= -\ev{\comm{\hat{x}(t)}{\hat{p}}^{2}}
\end{equation}
where, $\beta=\dfrac{1}{k_B T}$, also known as ``the coldness function", $\hat{x}$ and $\hat{p}$ are respectively quantum mechanical operators for position and momentum.

Using the classical-quantum correspondence\cite{COTLER2018318, PhysRevD.103.023533}, $ \dfrac{[,]}{i\hbar}\rightarrow \{,\}_{poisson}$, one can show its subtle relation with the classical Lyapunov exponent as follows,

\begin{equation}\label{eq3}
	\begin{array}{r@{}l}
		\frac{1}{i\hbar}\expval{[x(t),p(0)]} &\rightarrow \{x(t),p(0)\}=\pdv{x(t)}{x(0)}\sim e^{\lambda_L t}\\
		-\frac{1}{\hbar^2}\expval{[x(t),p(0)]^2} &\rightarrow \{x(t),p(0)\}^2=\qty(\pdv{x(t)}{x(0)})^2\sim e^{2\lambda_L t}
	\end{array}
\end{equation}

Thus, from the above semi-classical connection, OTOC quantifies the sensitivity of time evolutions in both chaotic and non-chaotic quantum systems to their initial conditions. In non-chaotic systems, fluctuations exhibit periodic, aperiodic, or irregular behaviour, whereas, in the case of chaotic quantum systems, the random time disordering phenomenon is characterized by exponential growth\cite{sym13010044, García-Mata:2023, PhysRevLett.122.024101, PhysRevB.100.035112, PhysRevB.105.224307}. Here, we assume that the classical-quantum correspondence can break down after the Ehrenfest time, making it more challenging to detect exponential developments in quantum systems compared to classical ones\cite{Maldacena2016}.

For a natural Hamiltonian with the form,
\begin{equation}\label{eq4}
	\mathcal{H}= \sum_{i=1}^{N}\dfrac{p_i^2}{2}+\mathcal{U}(x_{1},x_{2},...,x_{N}),
\end{equation}
and partition function,
\begin{eqnarray}\label{eq5}
	Z(\beta)=\sum{e^{-\beta\mathcal{H}}}=\sum_{m}{e^{-\beta E_m}}
\end{eqnarray}
eq.(\ref{eq2}) takes the form \cite{Hashimoto2017, Hashimoto2020, sym13010044, PhysRevLett.128.150601, Tetsuya_Akutagawa2020},
\begin{equation}{\label{eq6}}
\mathcal{C}_{\beta}(t)=\dfrac{1}{Z(\beta)} \sum _{m}e^{-\beta E_{m}} c_{m}(t) 
\end{equation}
where, $\mathcal{H}\ket{m}=E_m\ket{m}$ and
\begin{widetext}
\begin{equation}{\label{eq7}}
c_m(t)=\dfrac{1}{4} \sum _{k,l,r}^{trunc} x_{ml}~ x_{lk}~ x_{rm}~ x_{kr} glb( E_{rk}~E_{lk} e^{ i t E_{rl}} +E_{mr}~E_{ml} e^{- i t E_{rl}}  -E_{rk}~E_{ml}e^{it(E_{rm}+E_{lk})} - E_{mr}~E_{lk}e^{-it(E_{rm}+E_{lk})}glb)
\end{equation}
\end{widetext}

We refer $ c_m(t) $, in Eq.(\ref{eq7}), for a fixed energy eigenstate as \textit{Microcanonical OTOC}\footnote{OTOCs within the microcanonical ensemble, where the energy of the system is fixed, offers valuable insights into how information gets scrambled in quantum systems.} and $ \mathcal{C}_{\beta}(t) $, in Eq.(\ref{eq6}), as a \textit{Thermal OTOC}\footnote{The thermal ensemble, also known as the canonical ensemble, describes a system at a specific temperature. In contrast to the microcanonical ensemble, in the canonical or thermal OTOC calculation the temperature is fixed.}. Here, $E_{nm} = E_n-E_m$, $x_{nm} = \matrixel{n}{\hat{x}}{m}$ and $p_{nm} = \matrixel{n}{\hat{p}}{m}$. Here, $ p_{mn} =\dfrac{i}{2}E_{mn}x_{mn} $ Ref.(Appendix \ref{Appendix-a}, Eq.(\ref{A3})).
 
\section{\label{sc3}Models}
The idea behind the choice of our models rest on the fact that a small symmetry-breaking perturbation term in a Hamiltonian disrupts the underlying symmetries of certain classical systems. This disruption triggers a cascade of non-linear resonances within the broken symmetry regions. The emergence of these resonances often leads to the onset of chaos. As these regions grow, the chaotic nature of the system intensifies, resulting in increased unpredictability and complexity\cite{masoumi2015symmetry, Reichl2021}. 

Let us consider, a Hamiltonian of the form of:
\begin{equation}\label{eq8}
	\mathcal{H}=\frac{-1}{2}\laplacian_{x}+V(x)+ ~\Lambda*\;x
\end{equation}
where,
\begin{subequations}
\begin{eqnarray}\label{eq9}
V(x)&\rightarrow& \text{Potential} \label{eq9a}\\
\Lambda*\;x &\rightarrow& \text{Perturbation}
\end{eqnarray}
\end{subequations}

We work in the natural units( $ \hbar = 1 $) and, without any loss of generality, we assume that the mass of the oscillator, $ m = 1 $. 

Double well potential models, that we have considered, have an \textit{Inverted Harmonic Oscillator} (IHO) \cite{Aquino2011EnergyEF, 10.21468/SciPostPhysCore.4.1.002,Gentilini2015, SUBRAMANYAN2021168470, Choudhury_2022, Xu_2021, PhysRevD.106.106015, WEI2022115692, Du_2022} term. The IHO, a physically realised system, exhibits an unstable point at (x=0, p=0) in phase space. When perturbed, the particle undergoes exponential acceleration away from this fixed point, resulting in divergent solutions in phase space. $ Model~(I) $~(\ref{sc3A}), and $ Model~(I-a) $~(\ref{sec3Ai} :~ A special case), each has a distinct form of lower bounds and hilltops. Furthermore, $ Model~(II) $~(\ref{sc3B}) employs the Harmonic Oscillator (HO) potential form to create a triple-well configuration.

\subsection{\label{sc3A}$ Model~(I) $: Double Well}

For $ Model~(I) $, $ V(x) $ is in the form of a non-linear potential function:
\begin{subequations}
\begin{eqnarray}\label{eq10}
V(x)&=&a_0 ~\hat{x}^4-a_1 ~\hat{x}^2 \label{eq10a}
\\
~\Lambda \ &=& \qty(\sigma~\sqrt{\abs{\frac{a_0}{2 ~a_1}}}),\label{eq10b}
\end{eqnarray}
\end{subequations}
where, $ a_0 $, $ a_1 $ and $\sigma$ are known as stabilisation, destabilisation\footnote{$ a_0 $ determines the width of the wells and $ a_1 $ determines the curvature of the unstable top of the hill.} and asymmetry parameter, respectively. Note that the term $ x^4 $ provides a lower bound to the potential\footnote{Without the $ x^4 $ term in the potential, the system would not have any ground state and defining temperature ($ T $) would have been impossible.}, and the lowest bound is taken to be zero for all the models. The parameter $ \Lambda $ stands for the perturbation strength.

It is evident in this constructed system that the non-degenerate energy doublets have a splitting $ \Delta = \sigma$, i.e. splitting between levels remains close to $ \sigma $ for all doublets lying below the barrier top. The initial inspiration for devising the model is taken from \cite{https://doi.org/10.48550/arxiv.1107.0554, 10.1063/1.464760, BRIZARD2017351, M_Selg_2000}.

\subsubsection{\label{sc3Ai}Bifurcations and symmetry breaking}

\begin{figure}[H]
	\centering
	\subfigure[\label{fig1a} ]{\includegraphics[width=0.35\linewidth]{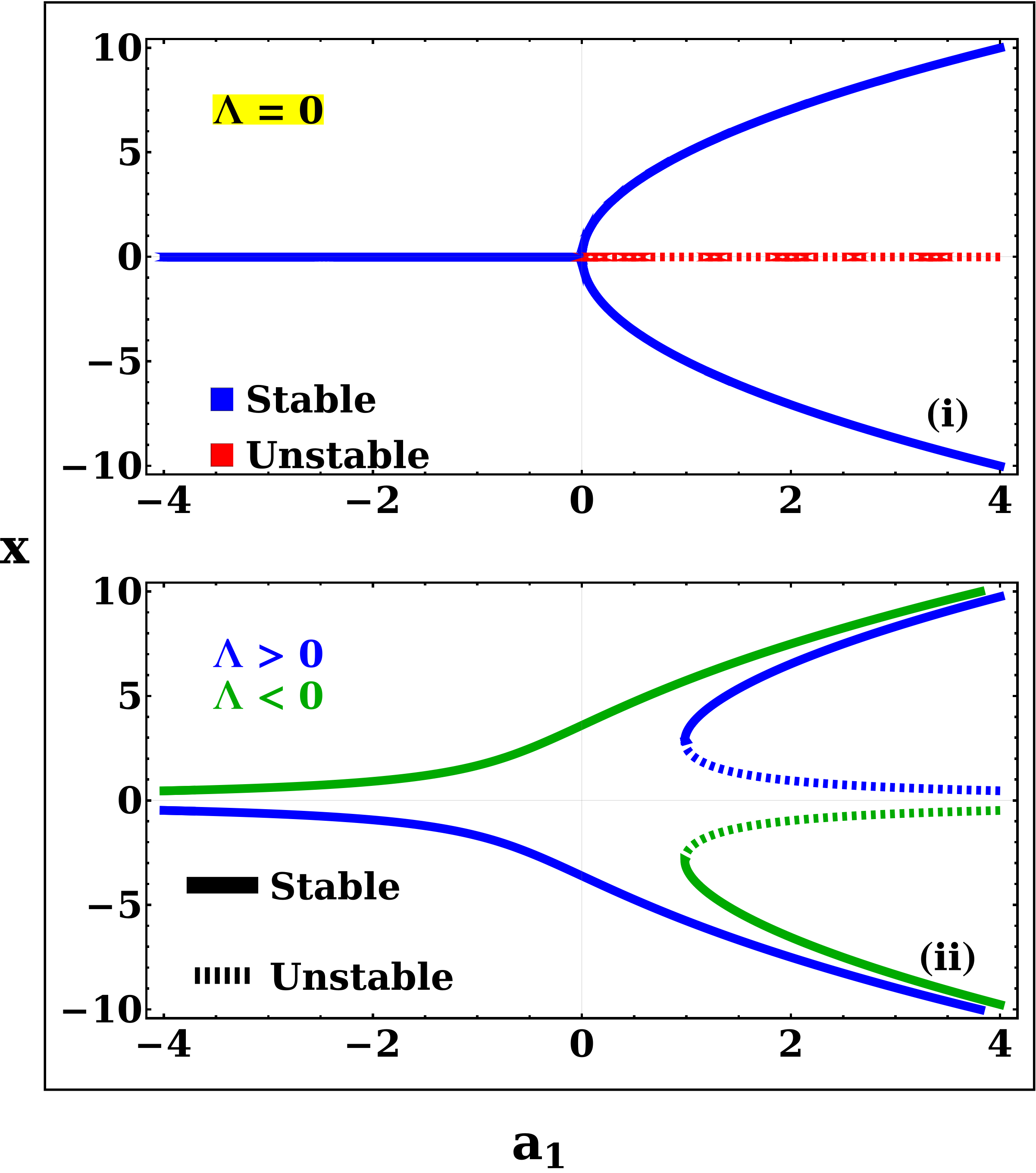}}
\hfill
	\subfigure[\label{fig1b}]{\includegraphics[width=0.61\linewidth]{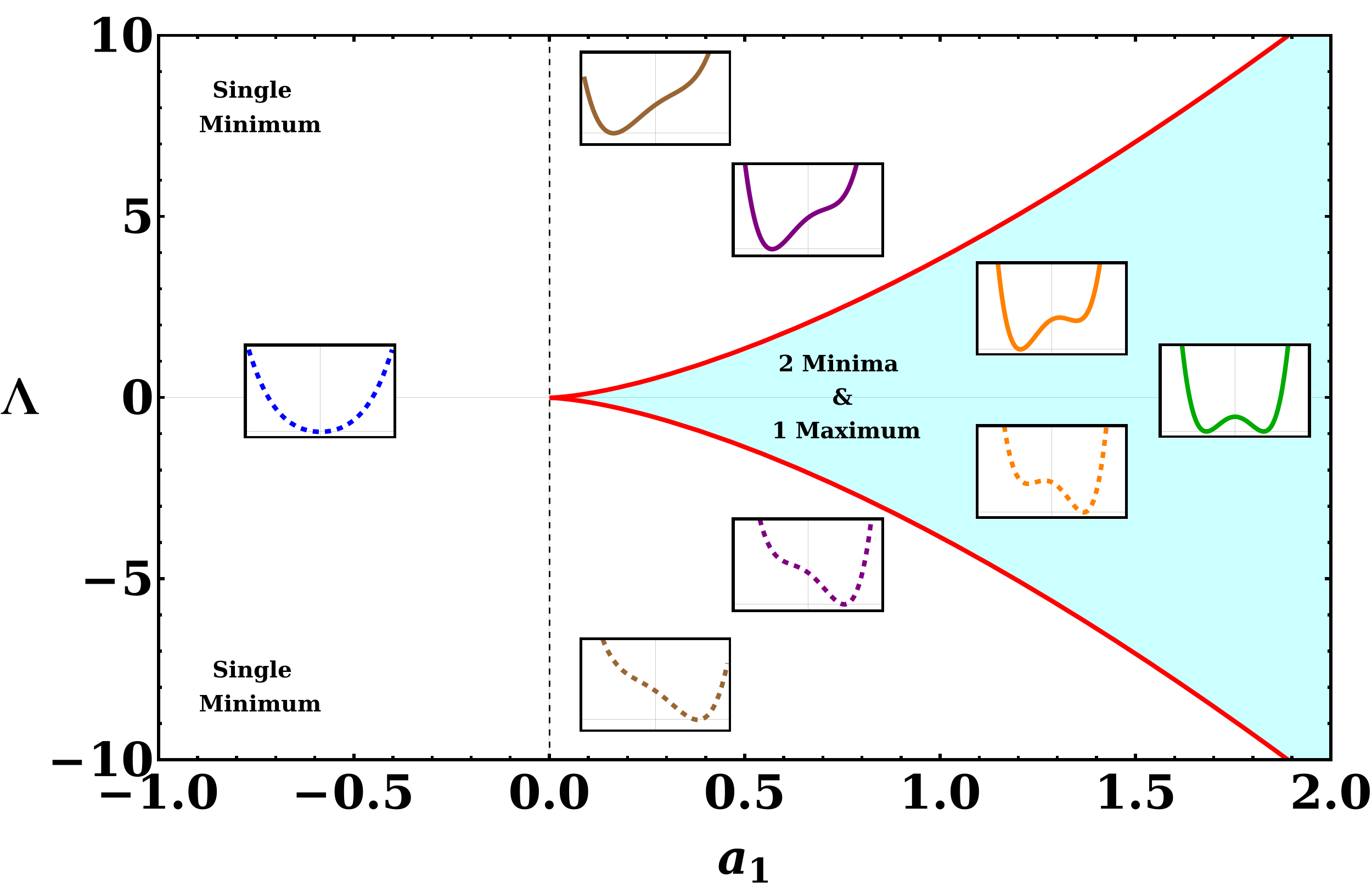}}	
	\caption{\label{fig1}Bifurcation and stability analysis graphs for $ Model~(I) $. $ (a) $ A saddle-node bifurcation occurs at $ a_1 = \dfrac{3}{2} a_{0}^{1/3} \Lambda^{2/3} $. $ (a(i)) $ For $\Lambda = 0$ this reduces to the supercritical pitchfork; $ (a(ii)) $ for finite $ \Lambda $ the pitchfork is deformed and even changed topologically. $ (b) $ Existence and stability domains of fixed points (or phase diagram) for the imperfect bifurcation in the $ (a_{1},~ \Lambda) $ plane are discussed. Here, the shaded region shows the domain where the system has three fixed points, and the plain white region shows the domain with one fixed point.}
\end{figure}

The pitchfork bifurcation is commonly encountered in systems in which there is an overall parity symmetry $ (x \rightarrow -x) $\cite{PhysRevE.74.056608, strogatz:2000, GAETA19901, PhysRevE.98.042209}, which has fixed points at $ x^* = 0 $ and $ x^* = \pm\sqrt{\abs{\frac{a_1}{2 a_0}}} $. The imperfect bifurcation occurs when a symmetry-breaking term is added to the pitchfork. Here, the constant $\Lambda$ breaks the parity symmetry if $ (x \rightarrow -x) $. The bifurcation diagrams are shown in Fig. (\ref{fig1a}), where the Fig. (1 (a-i)) shows pitchfork bifurcation when $\Lambda=0$ and the Fig. (1 (a-ii)) shows the imperfect bifurcation when the $\Lambda\neq0$. The system’s solution space becomes asymmetrical because of the presence of finite $\Lambda$. The solutions now favour one direction over the other instead of being perfectly symmetric about the origin. This can lead to the emergence of new stable or unstable equilibrium points, depending on the specific magnitude of $\Lambda$. Consideration of stability analysis or fixed point analysis around bifurcation points is necessary in order to determine how $\Lambda$ affects the stability landscape.

The fixed point analysis in Fig. (\ref{fig1b}) shows that the system can exhibit one or three fixed points (one global minimum, or a local maximum and two global minima). We shall restrict ourselves to the cases where the system has three fixed points. Hence, as long as $ a_{1}>0 $, the system has a double well potential. Even though $\Lambda < 0$ gives the exact parity inverse of the potential as for $ \Lambda > 0  $, we shall restrict ourselves to the latter case where $\Lambda > 0$ for all our studies. Notice that this asymmetric double well potential becomes symmetric when the relation $ \Lambda= 4 a_0 a_1 $ is satisfied. 

We numerically solve the time-independent Schrödinger equation considering the Hamiltonian in  Eq.(\ref{eq8}) for the potential given in Eq.(\ref{eq10a}) in \texttt{Mathematica 12.0} using \texttt{NDEigensystem} built-in command and obtain the energy eigenvalues $ E_{n} $ and the wave functions $ \Psi_{n}(x) $. The results for $ Model~(I) $ with $\sigma=0$ is shown in Fig.(2(a)). Here, the eigenstates below the hilltop are almost degenerate even though the Hamiltonian and the potential are symmetric under space inversion\cite{RJWHodgson_1989, Gupta2002, sakurai_napolitano_2017}. Fig.(2(b), 2(c), and 2(d)) show the energy eigenvalues for $ \sigma $ having values $ 10.0 $, $ 30.0 $, and $ 70.0 $, respectively. 

\begin{figure}[hbt!]
	\centering
\includegraphics[width=0.8\linewidth]{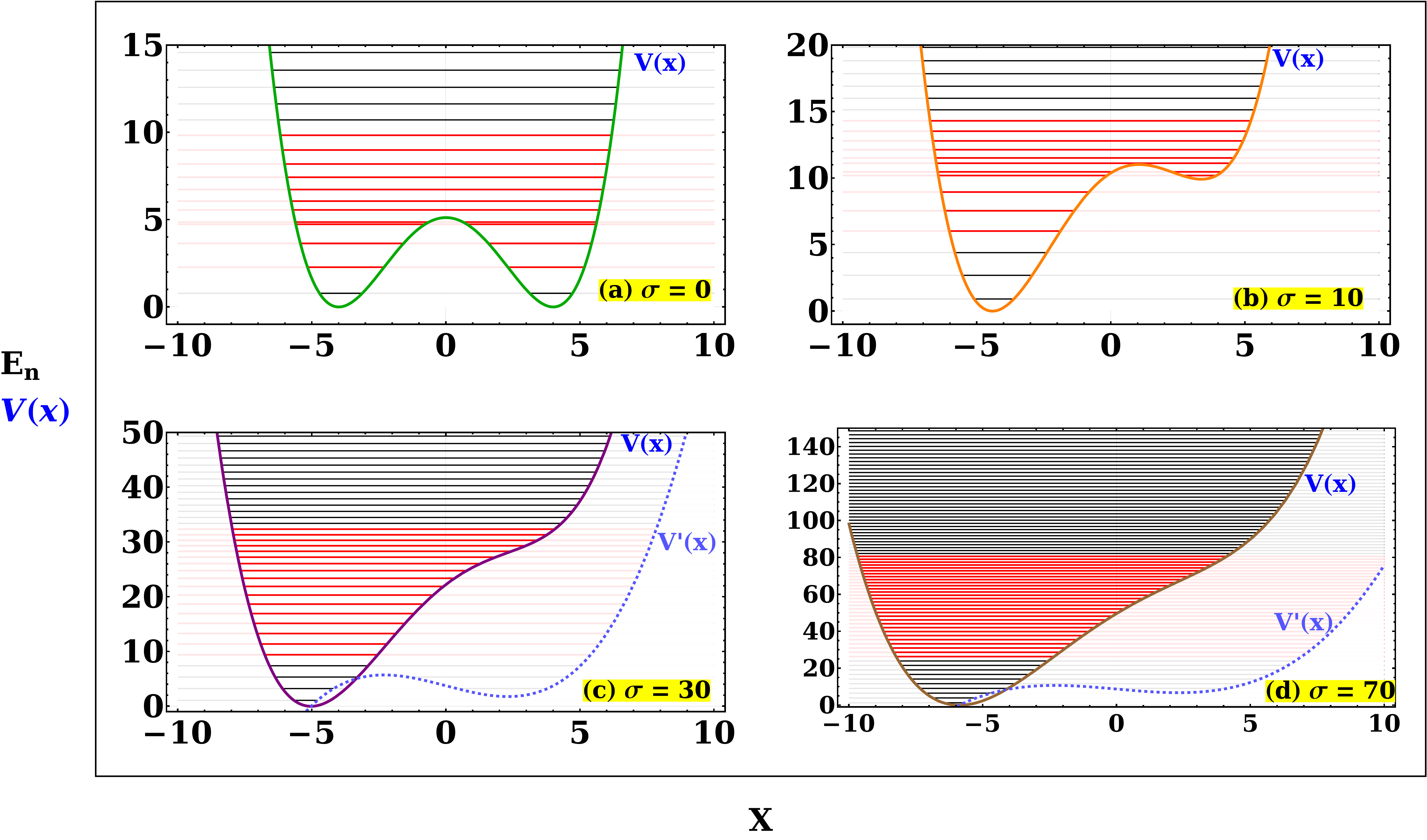}
	\caption{\label{fig2}Potential energy curves and energy eigenvalues for $ Model~(I) $ with 	$ a_0 = 0.02$, $a_1 = 0.64$ and $(a)~ \sigma=0.0 $ (two symmetric global minima), $(b)~ \sigma=10.0 $ (one global and one local minima), $(c)~ \sigma=30.0 $ (one global minima), $(d)~ \sigma=70.0 $ (one global minima). The energy levels below the top of the hill are almost degenerate, and the energy levels of the potentials in red colour show exponential growth in OTOC.}
\end{figure}

All our numerical observations on $ Model~ (I) $ and $ Model~ (I-a) $ are carried out by considering the width ($\Delta x$) to be 20, $ x $ ranging from $ -10 $ to $ 10 $. Below $ \Delta x = 15$, the energy separation between levels proliferates, causing the calculation of thermal OTOC very challenging. However, for $ Model~ (II) $, since it contains harmonic terms, so $\Delta x$ could range from $-\infty$ to $\infty$.

\subsubsection{\label{sc3Aii}OTOC analysis}
\begin{figure*}
	\centering
	\includegraphics[width=\linewidth]{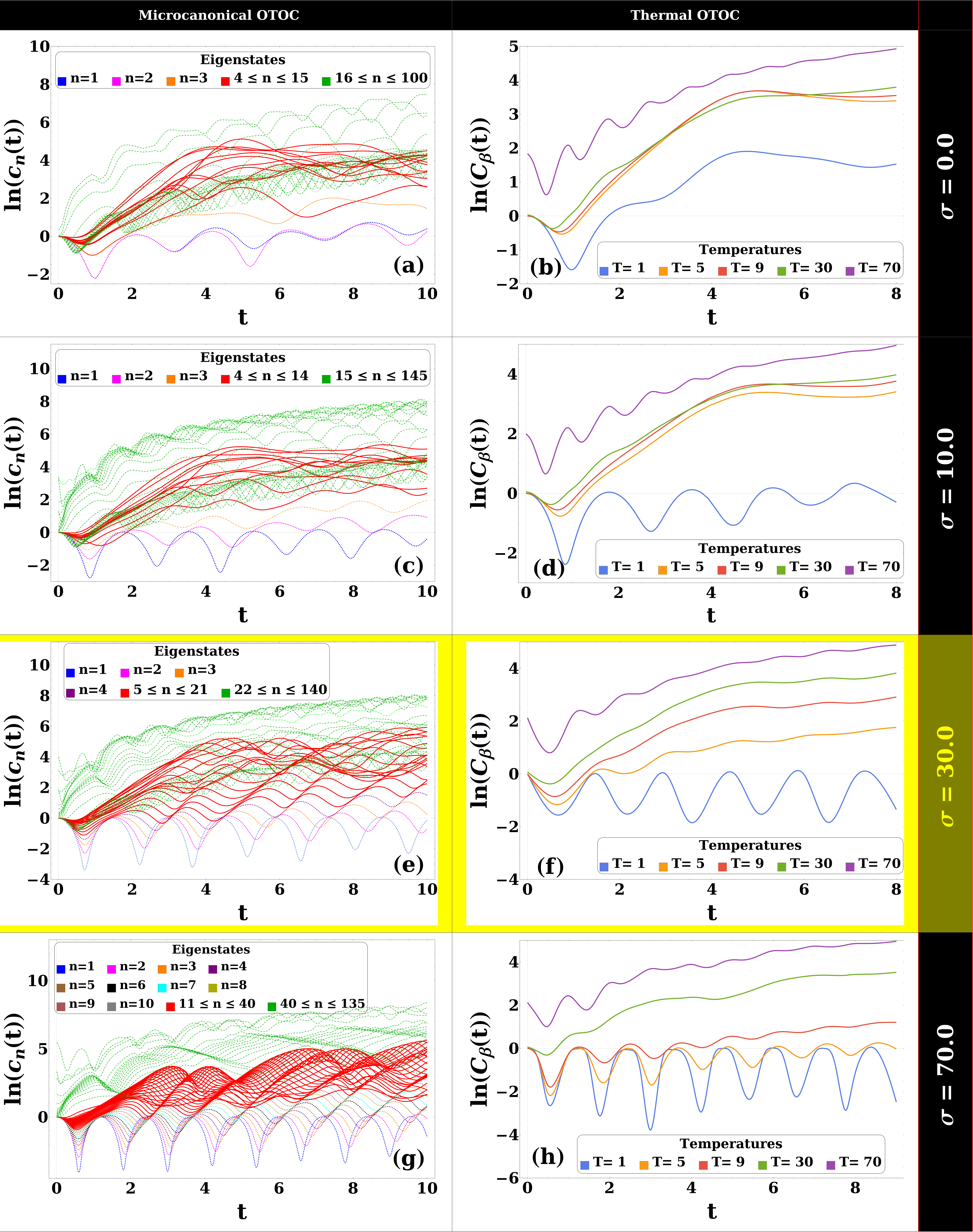}
	\caption{\label{fig3}Microcanonical and Thermal OTOCs for $ Model~(I) $ are in pairs in $ (a) \& (b) $, (c) \& (d), (e) \& (f), and (g) \& (h), respectively with $ \sigma=0.0 $, $ 10 $, $ 30 $, and $ 70 $. The exponential growth that occurred in the early times are illustrated in “red” colour for Microcanonical OTOC.}
\end{figure*}

In Fig. (\ref{fig3}), the microcanonical and thermal OTOCs are presented in four consecutive rows, with $\sigma=0$, $10$, $ 30 $ and $ 70 $. The left column shows the graphs of the Microcanonical OTOCs, while the right column displays the corresponding Thermal OTOCs. Each row of graphs represents a specific value of $\sigma$. 

The microcanonical OTOC for $ \sigma=0$ display exponential growth within a small energy eigenfunction range (from $n = 4$ to $15$) around the local maximum of the potential at $ E= 5.12 $, as shown in Fig. (3(a)). Furthermore, we compute the thermal OTOCs for various values of the temperature as shown in Fig. (3(b)). The numerical findings show exponential growths as temperatures rise. The primary factor behind the exponential growth of thermal OTOCs is the prominent role played by states that exhibit microcanonical OTOC behaviours.

We also observe the exponential growth of OTOCs for $\sigma = 10.0$ around the summit of the potential function. It should be noted that as the asymmetry parameter increases, there is an expansion in the broken symmetric region, corresponding to deformed pitchfork bifurcation. This causes an increase in the number of states exhibiting exponential growth. These two outcomes ($ \sigma=0 $ \& $ 10 $) agree with the findings of \cite{PhysRevLett.124.140602, Hashimoto2020}. However, at $\sigma = 30.0$ (the local maximum disappears for $\sigma>15.7656$ or $ \Lambda > 1.9707 $ $(= \dfrac{2}{3}\sqrt{\dfrac{2 a_{1}^{3}}{3 a_{0}}}) $). In the absence of a local maximum, both microcanonical and thermal OTOCs display sustained exponential growth (Fig. (3(e)) and 3(f)) within the Ehrenfest time scale. Here, the number of energy levels that show exponential growth has significantly increased from $ n =5 $ to $ 21 $ with the further widening of the broken symmetric region. At $ \sigma = 70.0 $ (Fig. (2(d))), without a local maximum, the deformation on the potential curve further smoothens. Here, a broader energy levels spectrum (from $ n = 11 $ to $ 40 $) exhibit exponential growth in microcanonical OTOC (Fig.(3(g))), albeit with the shortest growth duration. However, in this case, the thermal OTOC shows no such growth (Fig. (3(h))).

In each of these cases, the OTOCs for higher energy eigenstates initially show brief periods of growth followed by irregular oscillations or fluctuations. The brevity of this growth phase makes it challenging to discern its precise nature; whether it follows a linear, exponential, or polynomial pattern. Nevertheless, this pattern suggests that the influence causing exponential growth in the lower energy eigenstates diminishes as we move to higher energy levels. The irregular fluctuations indicate a transition to a more regular behaviour.

\begin{figure}[H]
	\centering
\subfigure[\label{fig4a}]{\includegraphics[width=0.51\linewidth]{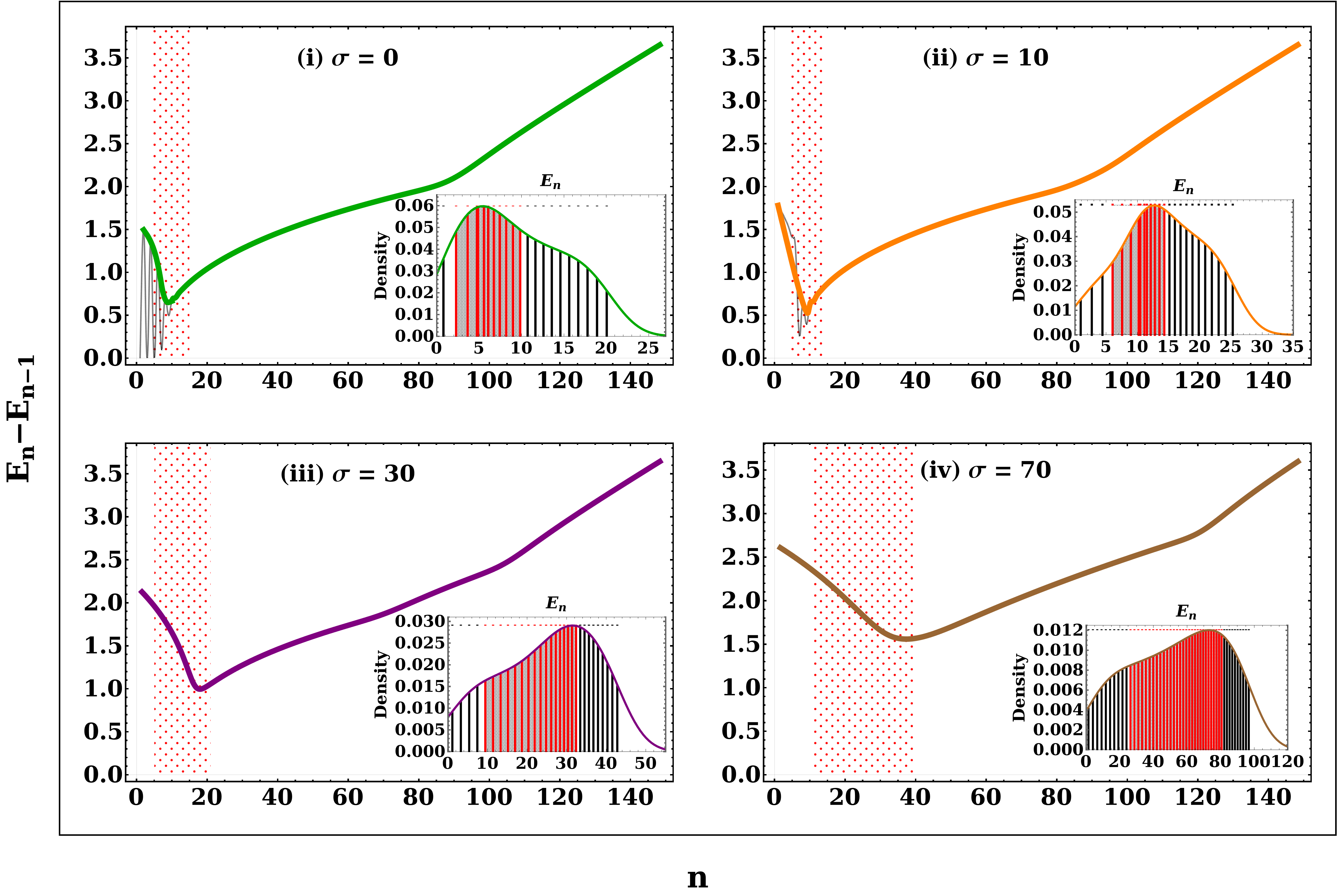}}
	\hfill
\subfigure[\label{fig4b} ]{\includegraphics[width=0.45\linewidth]{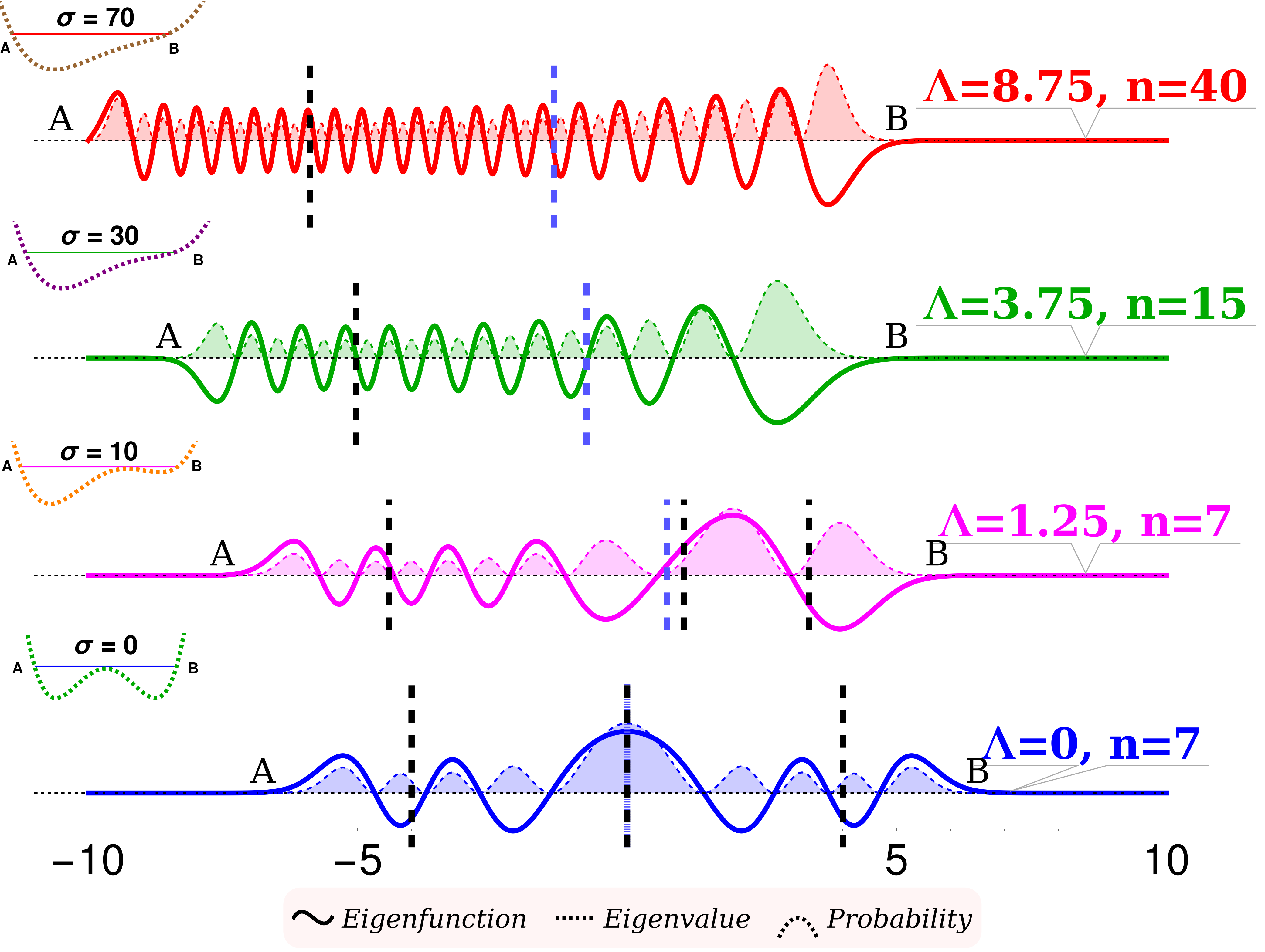}}	
	\caption{\label{fig4} (a) Difference in successive energy eigenvalues as a function of quantum number $ n $ for different values of $\sigma$. Inset plots show density of states. $ (a(i)) $ $ \sigma=0$, $ (a(ii)) $ $ \sigma=10$, $ (a(iii)) $ $ \sigma=30$, and $ (a(iv)) $ $ \sigma=70$. On the difference curve, the decreasing slope region is shaded in red. $ (b) $ Stretching of eigenstates (solid line) along the width of the potential and spreading of probability distribution (dashed line) for different $ \sigma $ values. The spreading and amplification of eigenstates is more for the states that lay in the red shaded regions of Fig (\ref{fig4a}). Notice that the positions of $ \expval{x}_{\Psi} $ (vertical black dashed lines) and the positions of equilibria (vertical blue dashed) indicate the symmetry-asymmetry of the potential.}
\end{figure}

Fig. (\ref{fig4a}) displays difference in successive energy eigenvalues. There are clear dips in all the curves (in red shaded region), which imply a higher density of eigenstates at a specific energy. The initial oscillatory behaviour (black dotted) for $\sigma=0$ \& $ 10 $, is due to the fact that eigenstates are almost doubly degenerate below the hill-top. As the asymmetry increases, so does the density of states. The figures also incorporate insets of the density of states graphs for the respective values of $\sigma$, offering an alternative viewpoint of the same analysis. The eigenstates belonging to these regions (red shaded regions) show exponential growths in OTOCs.

The graphs provide compelling evidence that the eigenstates, which show an exponential growth in the OTOCs, are tightly clustered together. This dip is near to the energy, where the slope of the potential is minimum at the classical turning point, having positive value. The nonlinear change of slope indicates deformation of asymmetry in the potential. In the case of a symmetric double well, this cluster happens to be located in close proximity to the hill-tops.

\subsubsection*{\label{sec3Ai}A special case: Double well potential with a plateau ($Model~(I-a)$)}

As a special case, we consider a double well potential with a plateau\cite{PhysRevD.106.106001, Garbaczewski_2022} whose lowest power is quartic rather than quadratic. This double-well potential is defined as below:
\begin{equation}\label{eq11}
	V(x)=a_0 ~\hat{x}^6-a_1 ~\hat{x}^4 
\end{equation}
where $ a_0~=\frac{1}{142} $, $ a_1~= 0.15$, $ x^4 $  produces a plateau instead of a hilltop and $ x^6 $ term gives a lower bound to the potential. As compared to $ Model~(I) $, this modified model ($ Model~(I-a) $) has a flat top (plateau).

\begin{figure}[H]
	\centering
\subfigure[\label{fig5a}]{\includegraphics[width=0.33\linewidth]{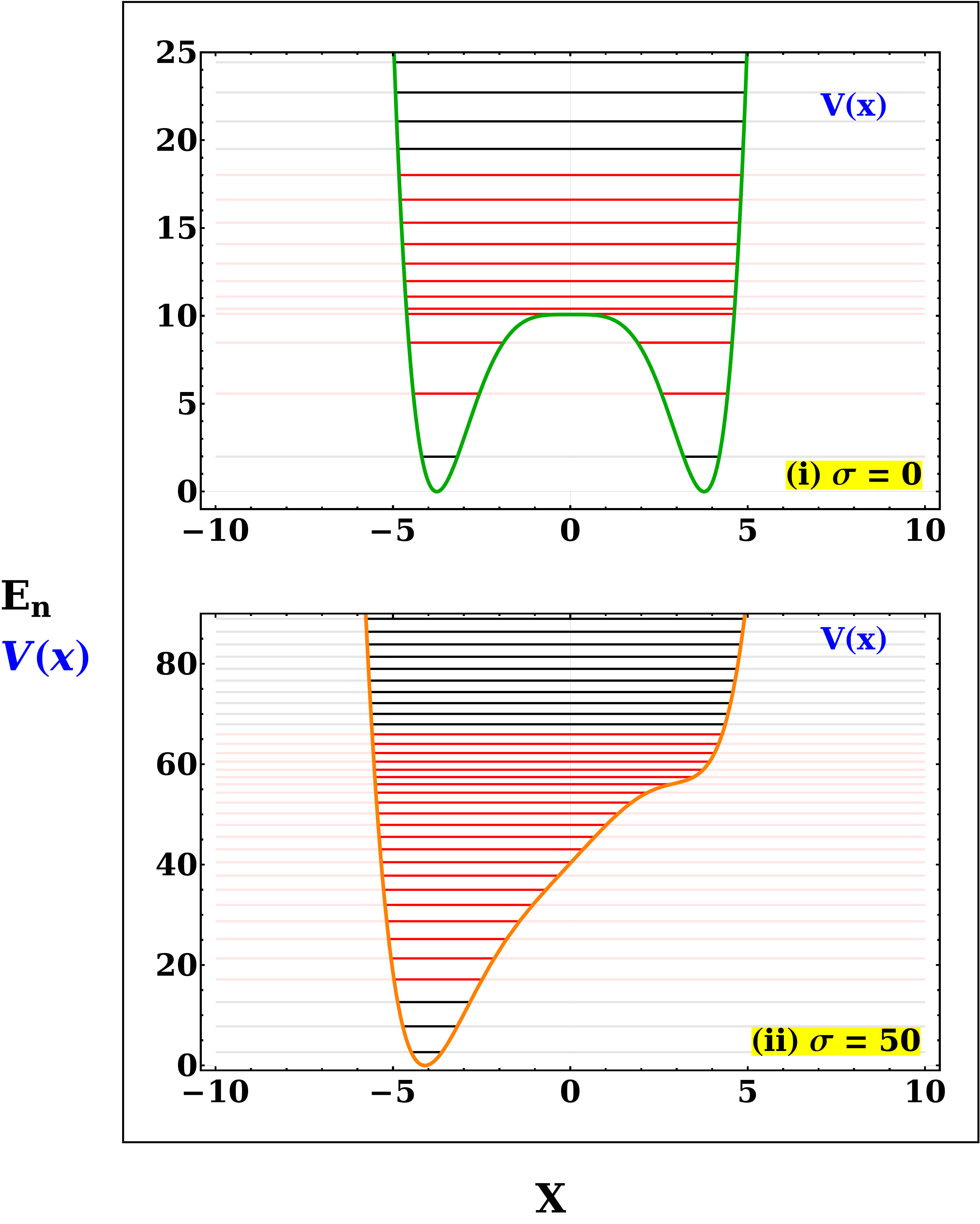}}
	\hfill
\subfigure[\label{fig5b}]{	\includegraphics[width=0.63\linewidth]{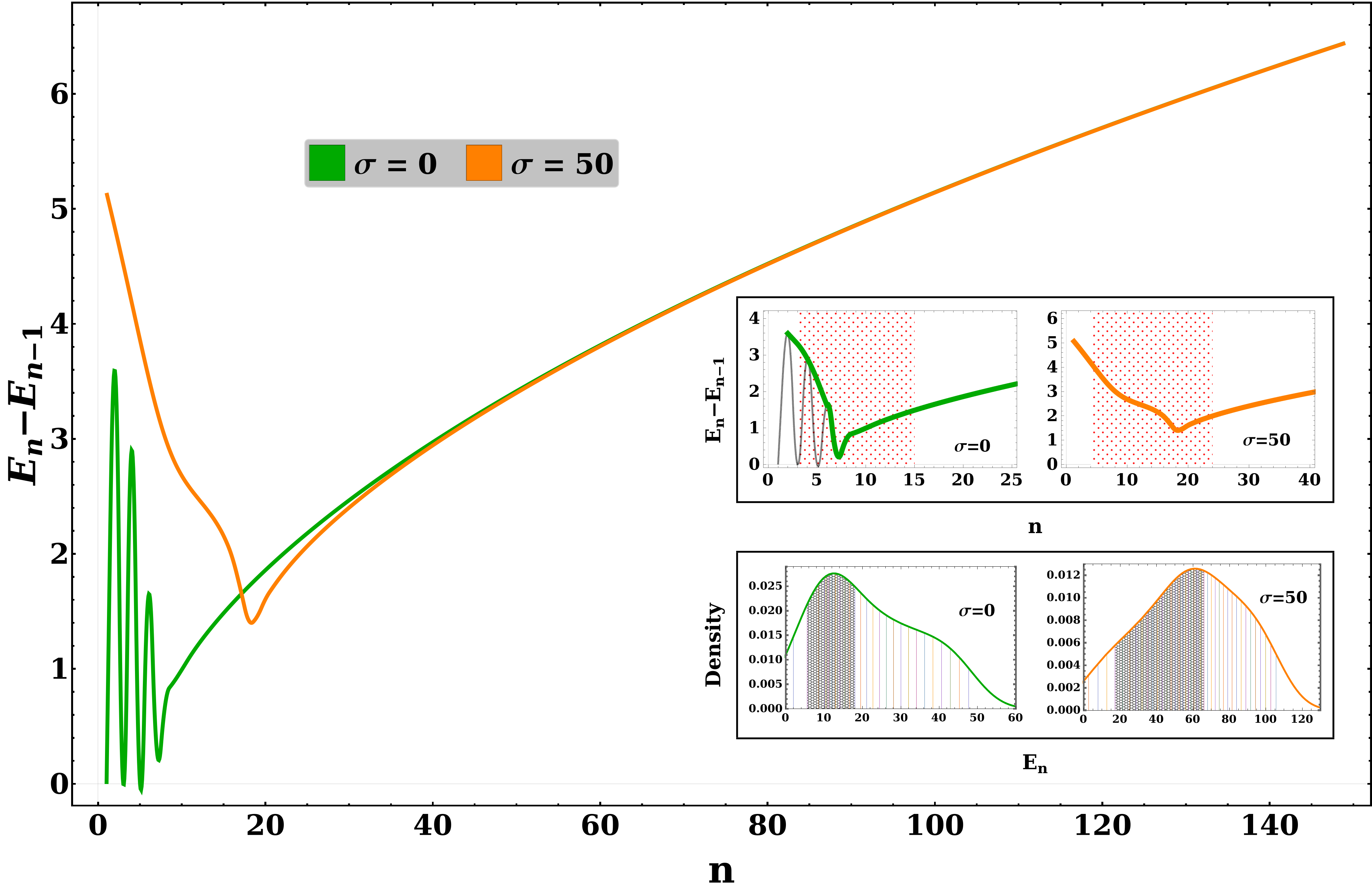}}
	\caption{\label{fig5}Potential energy curves and energy eigenvalues for $ Model~(I-a) $ with $(a-(i))~ \sigma=0.0 $, $(b-(i))~ \sigma=50.0 $. The energy levels below the top of the hill are almost degenerated, and the energy levels in red colour show exponential growth in the OTOC. (b) Difference in successive energy eigenvalues as a function of eigenstates for different values of $\sigma$. Inset plots show density of states. On the difference curve, the decreasing slope region is shaded in red.}
\end{figure}

Similar to $ Model~(I) $, here also $\Lambda$ breaks the parity symmetry leasing from pitch fork bifurcation to imperfect bifurcation. In Fig.(\ref{fig5}), the potential structure and corresponding energy eigenvalues for $\sigma = 0.0$, and $50.0$ are shown. Notably, with the increase in asymmetry strength, the plateau disappears entirely for $ \sigma > 38.9561 $ or $ \Lambda > 5.96857$ $(= \dfrac{16}{25}\sqrt{\dfrac{2 a_{1}^{5}}{5 a_{0}^{3}}}) $. 

Like in the previous case, we observe that the microcanonical OTOC shows exponential growth for energy levels near the plateau (from $ n = 3 $ to $ 15 $) for $\sigma=0.0$ (Fig. (6(a))). The effect of plateau is to bring the OTOC of these states closed to each other, forming a band and having almost the same growth rate. Accordingly, the exponential behaviour of thermal OTOC can also be seen in Fig. (6(b)) for higher value of temperatures. Similar behaviour persists for $\sigma>0$ (when asymmetry increases) including, $ \sigma=50 $, when the plateau is absent. Like in the previous model, exponential growth of OTOCs exist for shorter period than the Ehrenfest time for $ \sigma>>38.9561 $. In this case also there is a similar co-relation among OTOCs, energy eigenvalue distribution, the structure of eigenstates, and symmetry in the potential landscape $(Fig. (\ref{fig5b}))$.

\begin{figure}[H]
	\centering
	\includegraphics[width=0.925\linewidth]{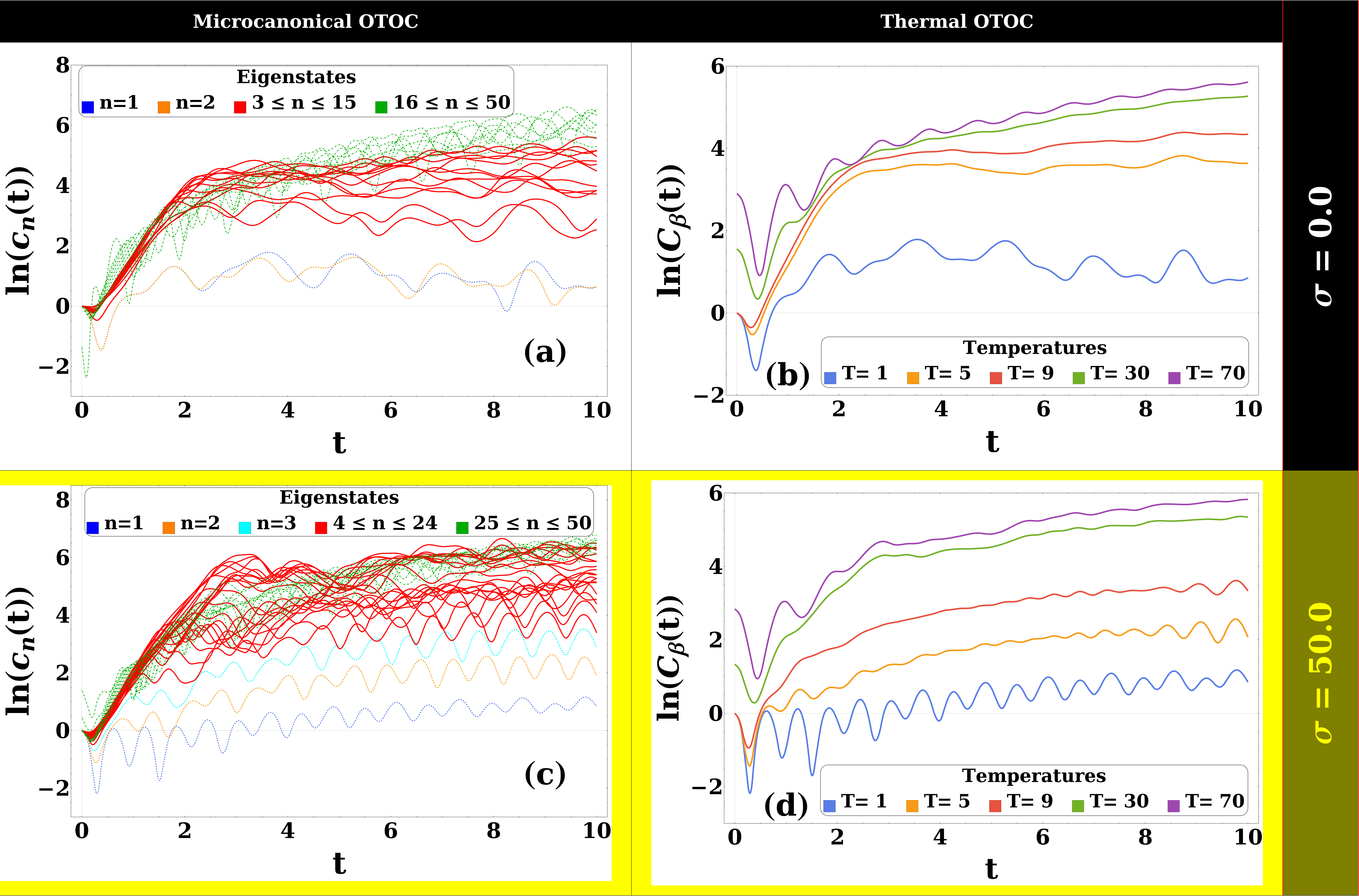}
	\caption{\label{fig6} (a), (b) shows microcanonical OTOC and (d), (e) shows thermal OTOC for $ Model~(I-a) $ with different values of $\sigma$.}
\end{figure}

\subsection{\label{sc3B}$ Model~(II) $ : Triple Well}

Triple-well potentials have applications in various areas of physics, such as the study of molecular dynamics, where they can model the behaviour of molecules in a triple-well potential energy landscape. They also have possible applications in quantum computing, where they can be used as a building block for quantum gates and algorithms.

This model is primarily inspired by the work on \cite{Aquino2011EnergyEF, PhysRevE.105.034204, PhysRevA.41.1218}. The potential in Eq. (\ref{eq9a}) for this case is
\begin{equation}\label{eq12}
	V(x)=a_1 ~x^2-a_0 ~x^4+x^6 
\end{equation}
where $a_0 = 10.95445$, $a_1 = 30.0 $.

\subsubsection{\label{sc3Bi}Bifurcation and symmetry analysis}

\begin{figure}[H]
	\centering
	\subfigure[\label{fig7a}]{\includegraphics[width=0.35\linewidth]{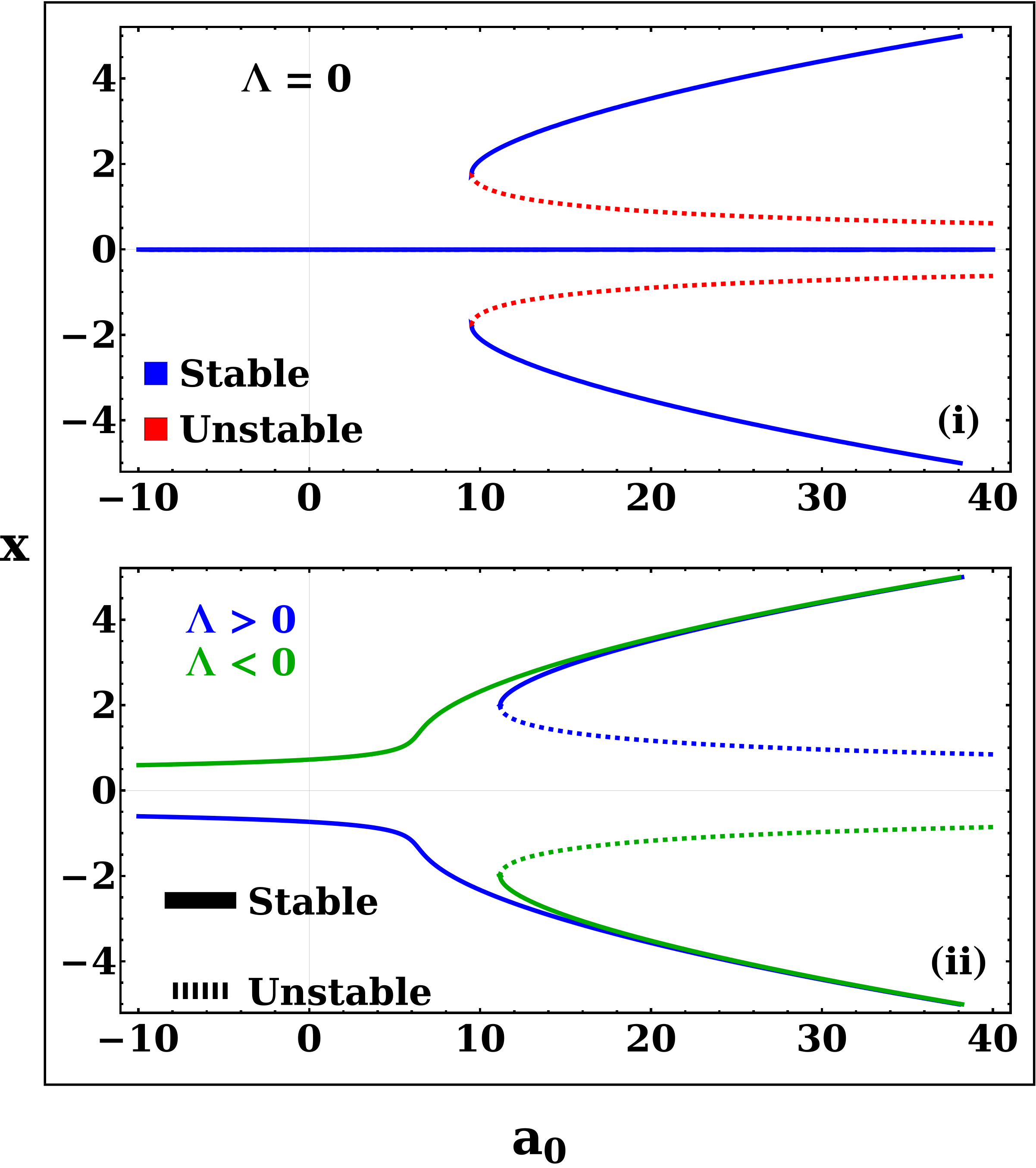}}
	\hfill	
	\subfigure[\label{fig7b}]{\includegraphics[width=0.61\linewidth]{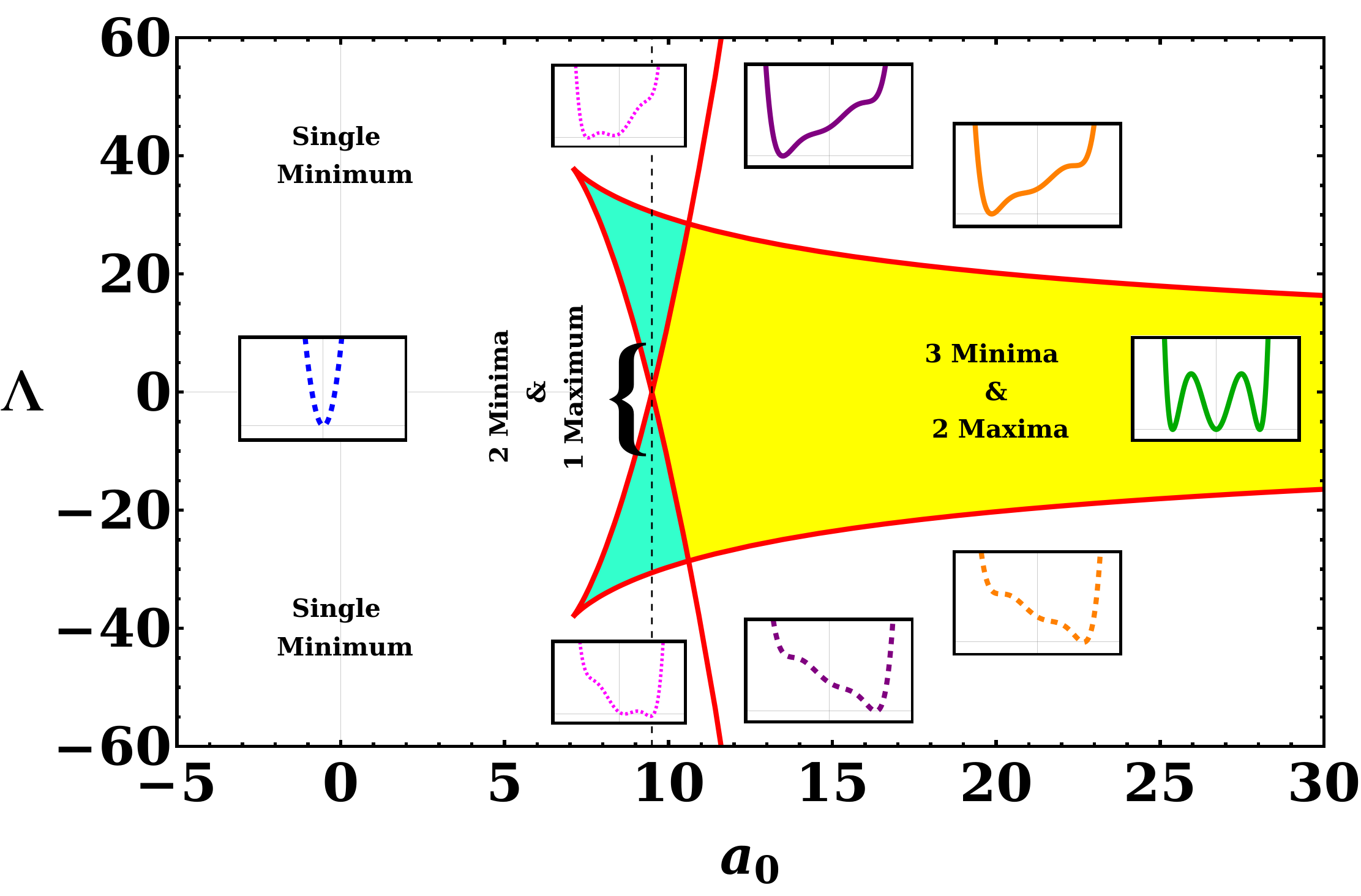}}
	\caption{\label{fig7}Bifurcation and stability analysis graphs for $ Model~(II) $. $ (a) $ Bifurcation Diagram. $ (b) $ Existence and stability domains of fixed points for the imperfect bifurcation in the $ (a_{0}, \Lambda) $ plane. Here, the yellow region shows the domain where the system has five fixed points, the cyan region contains the domain with two fixed points, and the plain white region shows the domain with one fixed point.}
\end{figure}

In Fig. (\ref{fig7a}), for $\Lambda = 0 $, the system always has a central minimum, along with two more minima of variable depths on either side of it depending on the value of $ a_{0} $\footnote{For $ 9.5<a_{0} $, we have a deformed single well potential. While for $ 9.5<a_{0}<10.95445 $, the central well is deeper, for $ a_{0}> 10.95445$, either-side wells are deeper than the central well.}. When $\Lambda$ has a finite value, the potential loss parity symmetry, ending up in an asymmetrical solution space. The solutions now favour one direction over the other instead of being symmetric about the origin (central minimum). This leads to the emergence of new stable or unstable equilibrium points. In order to assess the impact of $\Lambda$ on the stability landscape, stability analysis or fixed point analysis is performed around bifurcation points, as shown in Fig (\ref{fig7b}). Here, we have restricted ourselves to the cases where the system has five fixed points with $ \Lambda \ge 0  $, although $\Lambda < 0$ gives the exact parity inverse of the potential as for $ \Lambda > 0 $.

\begin{figure}[hbt!]
	\centering
\includegraphics[width=0.8\linewidth]{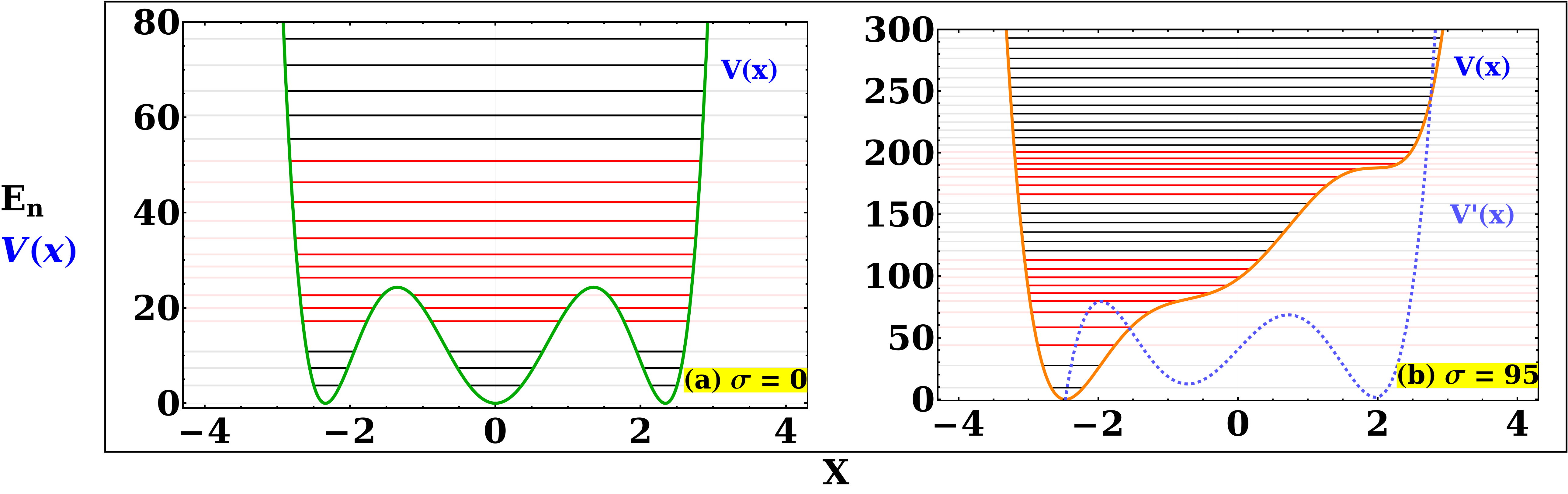}
	\caption{\label{fig8} Potential energy curves and energy eigenvalues for $ Model~(II)$ with different values of $ \sigma $. $ (a) $ For $ \sigma=0 $ (three global minima \& two local maxima), the energy levels below the top of the hill of left and right wells (except the middle) are almost degenerate. $ (b)  $ For $ \sigma= 95 $ (one global minimum). The energy levels in the red color exhibit exponential growth in OTOC.}
\end{figure}

Fig. (\ref{fig8}) shows the eigenvalue solution to the Schrödinger equation for the $ Model~ (II) $ potential with energy eigenvalue distributions under different values of $\sigma$. At $\sigma=0$, the eigenstates below the hilltops are non-degenerate. $\sigma > 90$ ensures the absence of a local maximum in the potential function. However, the energy level densities are found to be higher near two locations where the $ V^{\prime}(x) $ has two positive local minimum. The exponential growth of OTOCs is restricted to the eigenstates found in these regions. 

\subsubsection{\label{sc3Bii}OTOC analysis}

\begin{figure}[hbt!]
	\centering
\includegraphics[width=\linewidth]{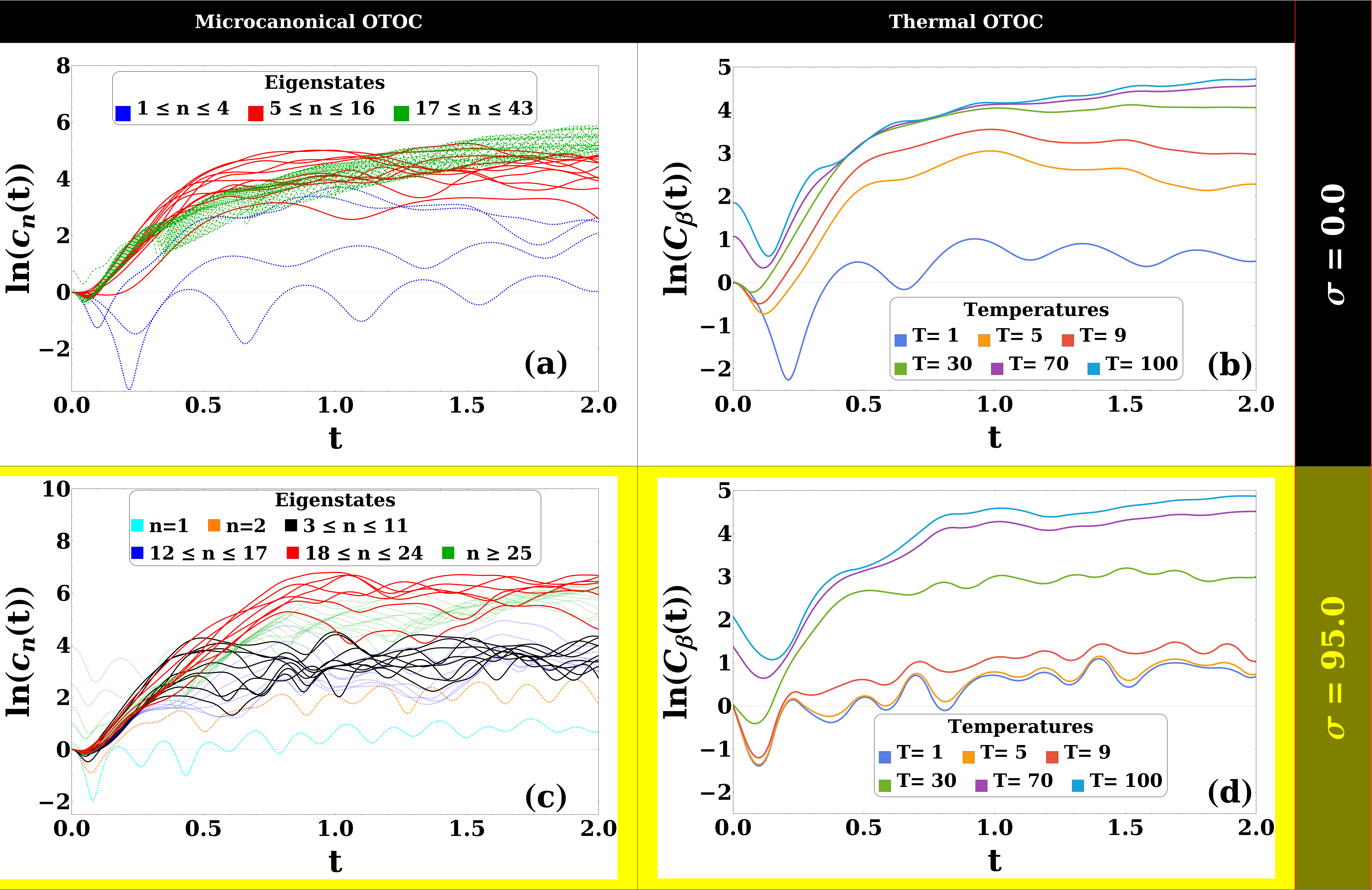}
	\caption{\label{fig9}Microcanonical and Thermal OTOCs for $ Model~(II) $ are shown in pairs in $ (a)$ \& $(b) $, and $(c)$ \& $(d)$, respectively with $ \sigma=0 $ and $ 95 $}
\end{figure}

In Figs. (9(a)), the microcanonical OTOC exhibits exponential growth around the hill-tops of the potential at $ E=24.3414 $ for $\sigma=0$. The energy eigenstates involved in this behaviour span from $n=5$ to $16$, with their energy ranging from $17$ to $51$. The exponential growth in thermal OTOC at higher temperatures is greatly influenced by these energy levels, as shown in Fig. (9(b)).

In Fig. (8(b)), at $\sigma= 95$, the microcanonical OTOC showcases exponential growth in two distinct energy level clusters (higher energy level densities) well separated along the potential landscape $(Fig. (9(c)))$. The first cluster spans from $n=3$ to $n=11$ (black lines), while the second spans from $n=18$ to $n=24$ (solid red lines). An important point is that the microcanonical OTOC for the second cluster (solid red lines) exhibits a marginally prolonged exponential growth with respect to time as compared to the first cluster (black lines). As shown in the Fig. (9(d)), the thermal OTOC also continues exhibit exponential growth for higher temperatures, but for a shorter time.

\begin{figure}[H]
	\centering
	\subfigure[\label{fig10a}]{\includegraphics[width=0.52\linewidth]{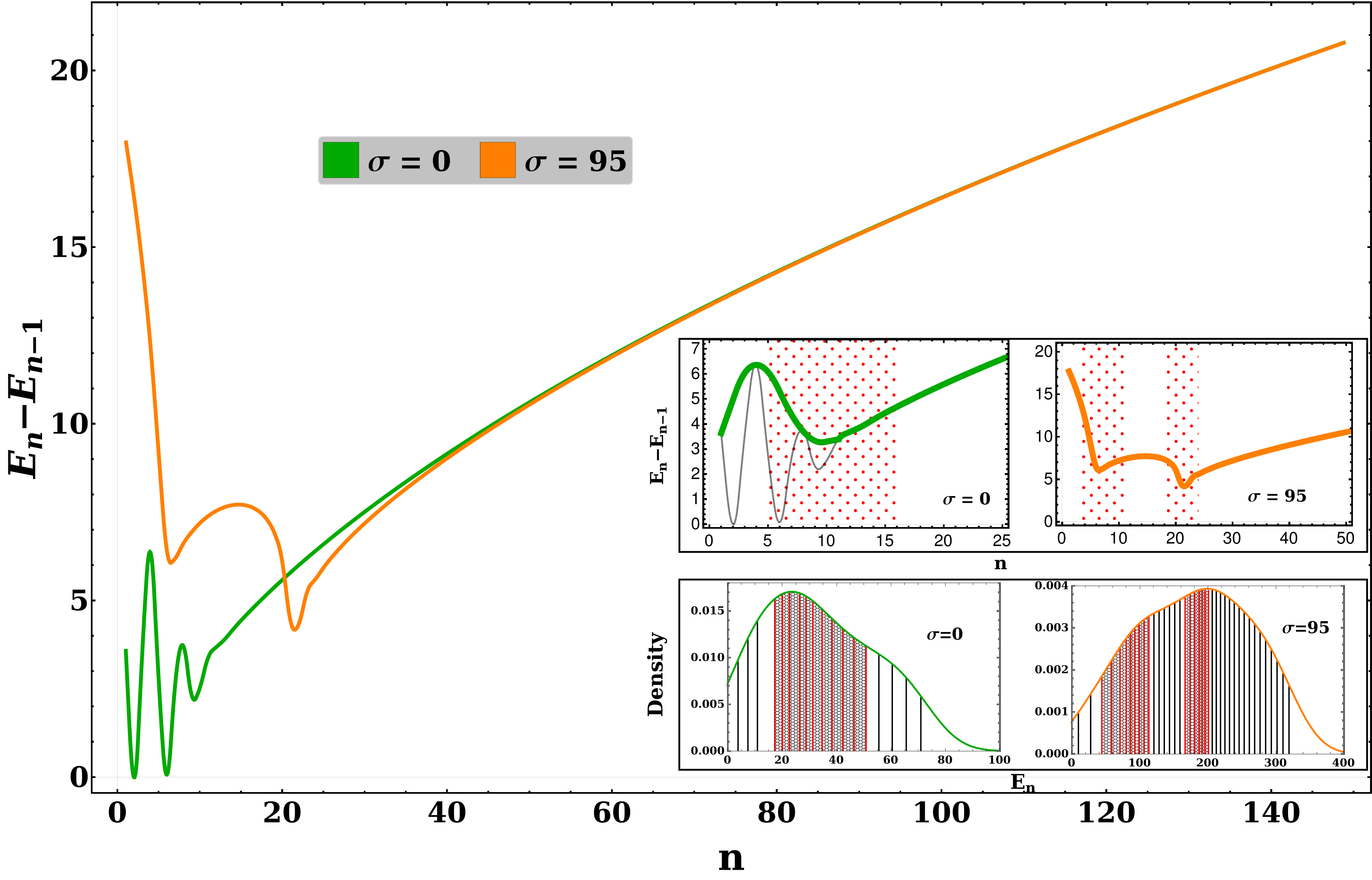}}
	\hfill
	\subfigure[\label{fig10b} ]{\includegraphics[width=0.44\linewidth]{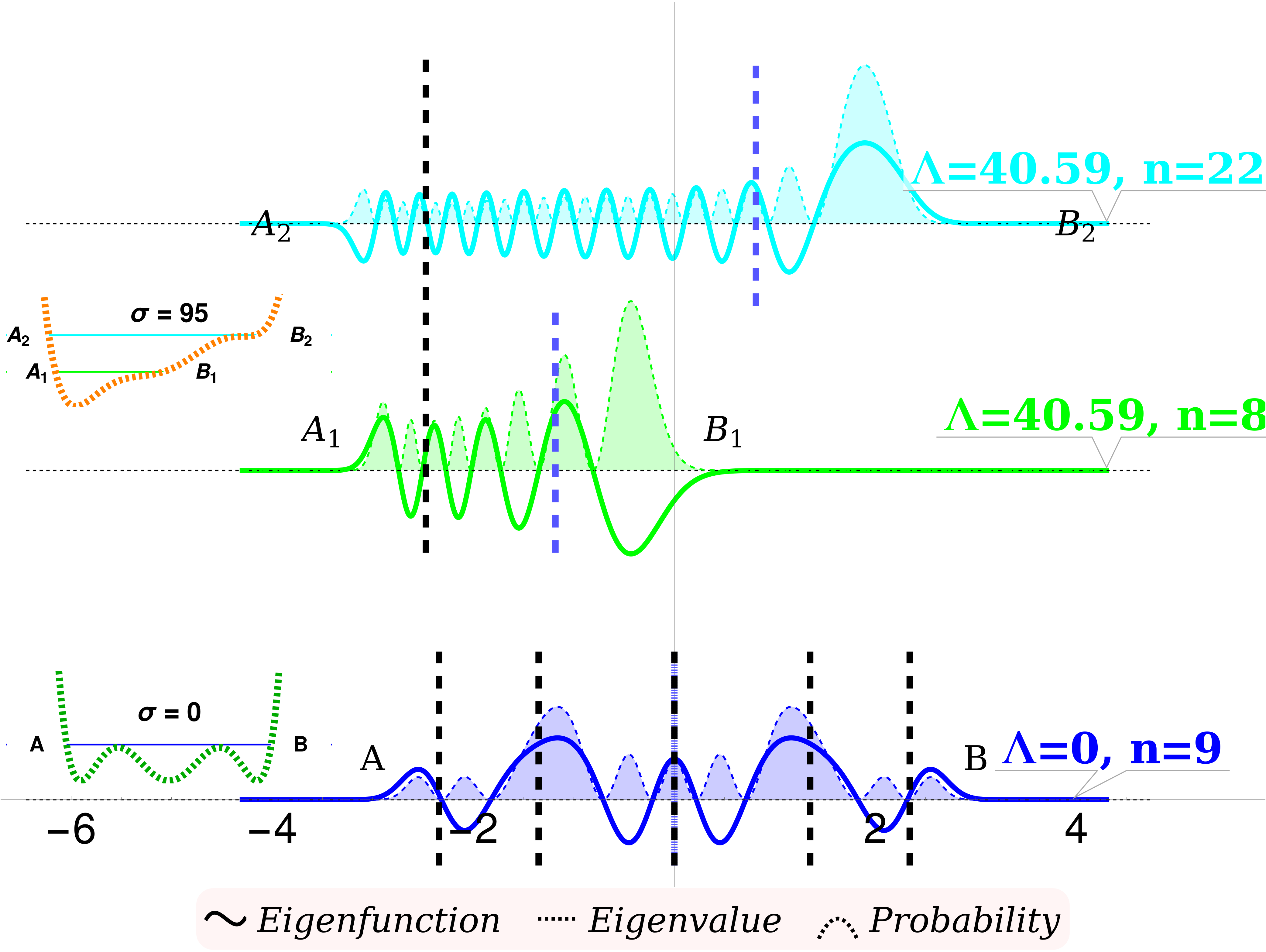}}
	\caption{\label{fig10}(a) Difference in successive energy eigenvalues as a function of eigenstates for different values of $\sigma$. Inset plots show density of states. On the difference curve, the decreasing slope region is shaded in red. (b) Stretching of eigenstates(solid line) along the width of the potential with spreading of probability distribution(dashed line) for different $ \sigma $ values. The spreading and amplification of eigenstates is more for the states that lay in the red shaded regions of Fig (4(a)). Notice that the positions of $ \expval{x}_{\Psi} $ (vertical black dashed lines) and the	positions of equilibria (vertical blue dashed) show the symmetry-asymmetry of the potential.}
\end{figure}

In the Fig. (\ref{fig10a}), we can see the successive differences in energy eigenvalues. The red shaded areas represent a higher density of states. For $\sigma=0$, following an initial oscillatory trend, a prominent decline emerges in the curve, representing a higher density of states. At $\sigma=95$, we can identify two distinct and well-separated dips in the curve. The first dip appears (in black shaded regions) to have a slightly shallower depth than the second (in red shaded regions), suggesting a higher density of energy eigenstates near the second dip. Observations suggest that there is a similar trend in the successive energy difference curve for higher values of $\sigma$. The Fig. (\ref{fig10a}) also incorporates insets of the density of states graphs for different $\sigma$ values and offering an alternative viewpoint of the same analysis. The eigenstates belonging to these regions (red and black shaded regions) show exponential growths in OTOCs.

These graphs offer clear evidence that the eigenstates, which show an exponential growth in the OTOCs, are tightly clustered together. This dip is near to the energy, where the slope of the potential is minimum at the turning point, having positive value. This observation aligns with the results of the $ Model~(I) $.

\subsection{\label{sc3C}Loschmidt Echo}

Loschmidt echo measures the revival occurring under an imperfect time-reversal procedure applied to a complex quantum system. This tool quantifies the sensitivity of quantum evolution to isospectral perturbations\footnote{No change in the spectrum of the unperturbed Hamiltonian}. Any spectral quantity will remain unaffected, but the Loschmidt echo will generally decay\cite{Chenu2018, Goussev:2012, GORIN200633, PhysRevLett.124.160603, PhysRevB.96.014202, PG_2021, PhysRevLett.124.030601, Ph.Jacquod_2003}.

Different decay behaviours in quantum fidelity are well-defined based on factors like perturbation strength, system complexity, dimension, and initial state. Reviews by Gorin \textit{et al.}\cite{GORIN200633} and Jacquod \textit{et al.} \cite{doi:10.1080/00018730902831009} discuss fidelity properties, decoherence, and irreversibility. Notable decay regimes include Lyapunov\cite{PhysRevLett.86.2490}, Fermi-Golden-Rule\cite{PhysRevLett.86.2490, PhysRevE.64.055203}, and perturbative\cite{ PhysRevE.64.055203, Nicholas_R_Cerruti_2003, Tomaz_Prosen_2002}, observed in both regular and chaotic systems\cite{PhysRevE.89.022915, PhysRevE.68.036216, Prosen_2003}. This contrasts the typical view of exponential decay as chaotic; to our knowledge, no theoretical model is currently able to quantitatively explain the exponential decay of the Loschmidt echo in integrable systems.

Mathematically, the Loschmidt echo is then defined as,
\begin{equation}\label{eq13}
	M(t)=\qty|A(t)|^2=\qty|\ev{e^{i\hat{\mathcal{H}}_2t/\hbar}e^{-i\hat{\mathcal{H}}_1t/\hbar}}{\Psi_0}|^2
\end{equation}
where $ A(t) $ is the Loschmidt amplitude. It quantifies the “distance” (in the Hilbert space) between the state $ e^{-i\hat{\mathcal{H}_1}t}\ket{\Psi_0} $, resulting from the initial state $ \ket{\Psi_0} $ in the course of evolution through a time $ t $ under the Hamiltonian $ \mathcal{H}_1 $, and the state $e^{-i\hat{\mathcal{H}_2}t}\ket{\Psi_0} $ obtained by evolving the same initial state through the same time $ t $, but under a slightly different perturbed Hamiltonian $\mathcal{H}_2$. The LE, by construction, equals unity for t = 0 and typically decays further in time.

Perfect recovery of $\ket{\Psi_0}$ would be achieved by choosing $\mathcal{H}_{2} = \mathcal{H}_{1}$, which leads to $ M(t) = 1 $, but this is an impossible task in realistic problems and $ M(t) $ is usually a decreasing function in $ t $. The notion of time reversal i.e. a backward time evolution from $t$ to $0$ under $\mathcal{H}_2$ is equivalent to the forward evolution between $t$ and $2t$ under the Hamiltonian $-\mathcal{H}_2$.

Now, to compute Eq.(\ref{eq14}), we follow the algorithm given by A. Peres in \cite{PhysRevA.30.1610}. It follows that 
	\begin{equation}\label{eq14}
		\begin{split}
			M(t)&=\qty|\ev{e^{i(\hat{\mathcal{H}}_2-\hat{\mathcal{H}}_1)t/\hbar}}{\Psi_0}|^2\\&
			=\qty|\ev{e^{i\Lambda \hat{x} t/\hbar}}{\Psi_0}|^2\\&
			=\qty|\ev{e^{i \hat{x} \tau}}{\Psi_0}|^2
		\end{split}
	\end{equation}
where $ \tau=\dfrac{\Lambda t}{\hbar} $ and we assume that $ 	\Lambda \hat{x} (=\hat{\mathcal{H}}_2-\hat{\mathcal{H}}_1)$ is classically small but quantum mechanically significant such that the perturbation does not change the topology of the trajectories but introduces a phase difference.

For regular systems, $ M(t) $ must oscillates with a fairly large amplitude. Strictly speaking, $ M(t) $ is almost periodic. On the other hand, if $ \mathcal{H}_1 $ is chaotic, $ M(t) $ is small. Another fact is that the initial decay rate of $ M(t) $ for regular systems is fairly same as for chaotic systems. It is in fact completely independent of Hamiltonian.

\begin{figure}[H]
	\centering
\includegraphics[width=0.7\linewidth]{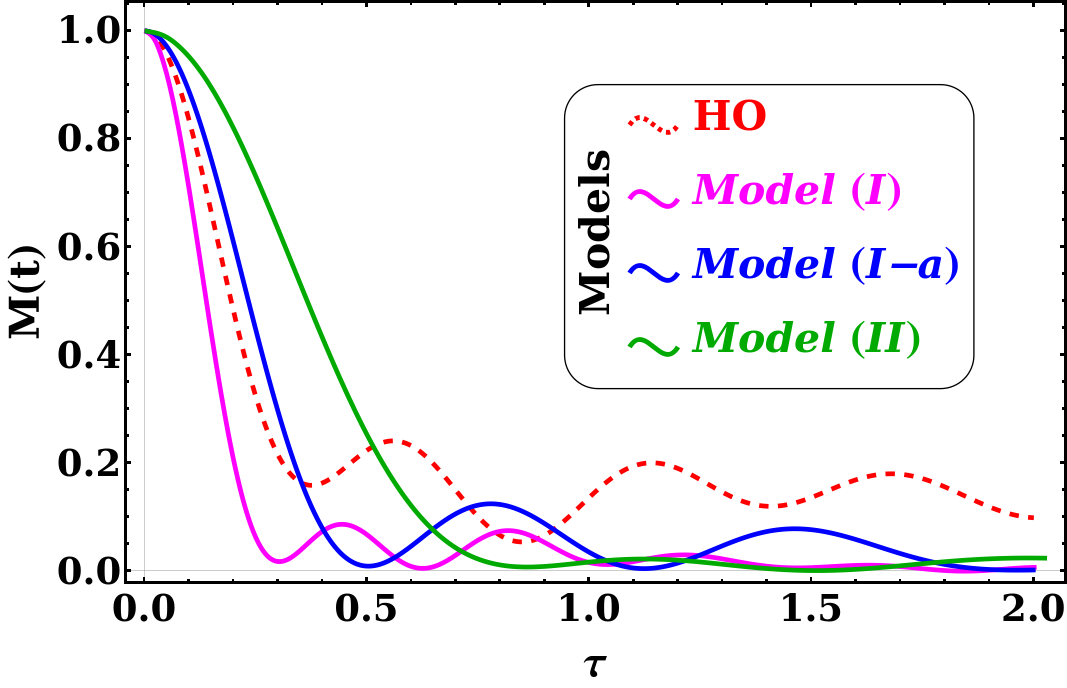}
	\caption{\label{fig11}Loschmidt Echo of Harmonic Oscillator({\color{red}HO}), {\color{magenta}$ Model~(I) $}, {\color{blue}$ Model~(I-a)$}, and {\color{green!60!black}$ Model~(II) $}. }
\end{figure}

For varying $\Lambda$ values, our numerical investigation reveals an initial exponential decay followed by anticipated low-amplitude oscillations in the LE. Fig. (\ref{fig11}) illustrates the compilation of LEs encompassing the Harmonic Oscillator(HO), as well as $ Models~ (I) $, $ (II) $, and $ (I-a) $. In particular, after the initial decay, the LE displays significant amplitude fluctuations in the HO case. However, intriguingly, our investigated models demonstrate comparatively smaller fluctuations in their LE behaviour. This contrast in fluctuation amplitude between the HO and our models suggests the presence of distinctive dynamical behaviours, potentially indicating unconventional characteristics in our systems.

\section{\label{sc4}Discussion}

As discussed in the introduction, the primary focus in the work of \cite{Hashimoto2020, PhysRevA.104.043308, PhysRevE.101.010202, PhysRevLett.124.140602, PhysRevA.103.033304, PhysRevLett.122.101603, PhysRevD.106.106001, PhysRevE.103.L030201, PhysRevLett.125.014101}, is to observe the behaviour of OTOC near saddle points. Our detailed investigations, using models with varying perturbations, show that there is a connection between exponential growth in OTOC, energy eigenvalue distribution, the structure of eigenstates, and symmetry in the potential landscape.

The emergence of a maximum in a single well can also result from a symmetry-breaking perturbation (pitchfork bifurcation), provoking the system to transition into a lower symmetric state. Due to this symmetry breaking, the linear behaviour in $ V^{\prime} (x) $ becomes nonlinear and its value becomes zero at the bifurcation point. The nonlinear behaviour of $ V^{\prime}(x) $ quantifies the asymmetry in the potential. If we examine $ Model~(I) $, we discover that $V^{\prime}(x) $ has a singular positive minimum at $ x=0 $, for $ \sigma =0 $ (in fact here $V^{\prime}(x) =0$) , while progressively shifting towards positive $ x $ as $ \sigma $ increases. This implies the asymmetry in the potential spreads towards positive $ x $. For a much higher value of $ \sigma $, the minimum in  $ V^{\prime} (x) $ is indistinguishable, implying no further spreading of asymmetry or deformation in the potential.

Our detailed study of OTOC (Fig. (\ref{fig3})), energy level density (Fig. (\ref{fig4a})), spreading of eigenfunction and probability distribution (Fig. (\ref{fig4b})) for $ Model~ (I) $ shows a strong co-relation among all these quantities and with asymmetry in the potential. These figures show that exponential growth in OTOCs emerges for those energy levels which are closely packed. These energy levels are in the neighbourhood of the classical turning point of the potential where the slope is a positive minimum. The dip in the energy difference curve (Fig. (\ref{fig4a})) is at the energy, which is equal to the potential where a positive minimum of the slope occurs. The corresponding energy eigenfunctions, near the turning points, resonate with the shape of the potential and create an Airy function \cite{doi:10.1142/p345, abramowitz2012handbook} like shape. As a result these energy eigenstates and their probability distribution have maximum spread over this region. The clusters of eigenstates involved in this behaviour lag behind or reach up to the dip on the energy difference curve.

The presence of plateau provides positive minima in $ V^{\prime}(x) $ over a range of $ x $. This causes the exponential growth of OTOCs for closely packed energy levels to have the same growth rate (Fig. (\ref{fig6})). Given that symmetry breaking bifurcation is topologically equivalent to the quartic case, analogous phenomena manifest in the energy difference curve (Fig. (\ref{fig5b})) and eigenstate dynamics. 

In the case of $ Model~ (II) $, two maxima arise in a single well because of symmetry breaking perturbation, after which the system ends up in a lower symmetric state. Perturbation gives rise to two simultaneous positive minima in $ V^{\prime}(x) $. As with $ Model~ (I) $, these two minima in $ V^{\prime}(x) $ shift towards positive $ x $ as we increase the strength of linear perturbation. Due to asymmetry, one of the minima becomes local and the other global. The global minimum in $ V^{\prime}(x) $ is slightly sharper than the local. Our detailed study of OTOC (Fig. (\ref{fig9})), energy level density (Fig. (\ref{fig10a})), spreading of eigenfunction and probability distribution (Fig.(\ref{fig10b})) correspond to these two minima points in $ V^{\prime}(x) $. 

The dips in the energy difference curve (Fig. (\ref{fig4a})) is at energies, which are respectively equal to the turning points on the potential function where two positive minima of the slope occurs. The higher dip corresponds to the local minimum, whereas the lower dip corresponds to the global. The lower dip (high density) is due to sharper curvature at the global minimum. Similar to $ Model~(I) $, here also we observe the spreading of energy eigenfunctions and maximisation of probability distributions over these regions. The eigenstates belonging to the high density of states show longer time exponential growth as compared to the eigenstates belonging to low density of states. In each of these cases, the clusters of eigenstates that show exponential growth in OTOCs lag or reach up to the dip on the energy difference curve.

\begin{figure}[hbt!]
	\centering
\includegraphics[width=0.7\linewidth]{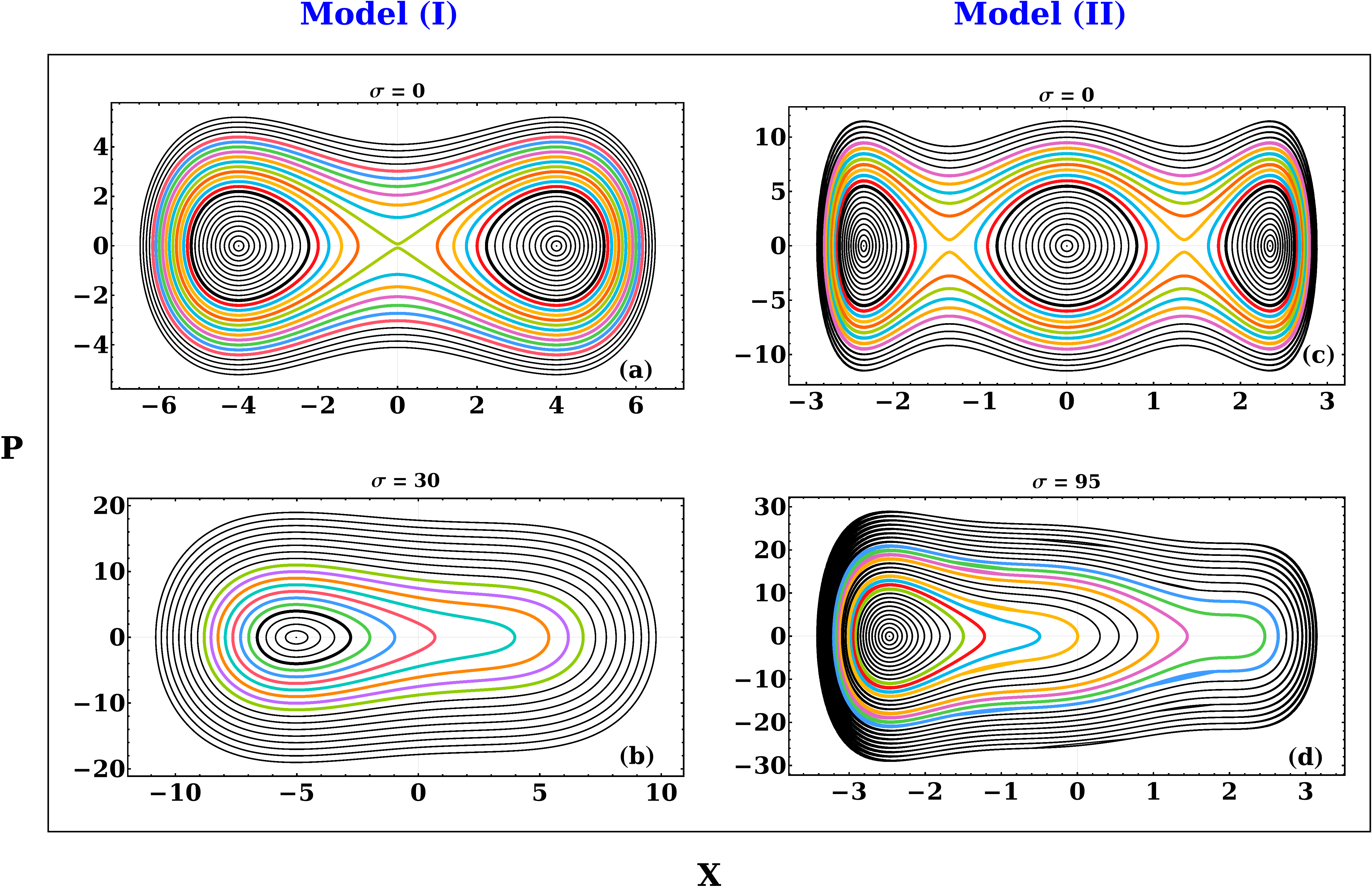}
	\caption{\label{fig13}Classical phase space diagram showing stretching of trajectories in symmetry and asymmetry hilltops.}
\end{figure}

There is a strong correlation of a quantum picture of OTOC, density of states, spreading of eigenstates and probabilities with the classical phase space. Fig. (\ref{fig13}) shows the phase space diagrams for $ Model~(I) $ and $ (II) $ Hamiltonians. The density of trajectories is low at the position where $ V^{\prime}(x) $ is a positive minimum. The OTOCs also show exponential growth when $ V^{\prime}(x) $ is a positive minimum.

It is a well known fact that there is a connection between OTOC and Loschmidt Echo \cite{PG_2021, bhattacharyya2022towards}. Exponential growth in OTOCs corresponds to small oscillation in Loschmidt Echo after the initial decay. Our calculations for perturbed systems without hilltops also show small amplitude oscillations in all the chosen models. This aligns with our observations of OTOCs. 

\section{\label{sc5}Conclusion}

Several earlier researches have shown that for the potentials having a hilltop, OTOCs show exponential growth for the eigenstates near the hilltop. The reason for such behaviour is attributed to the instability due to the hilltop. Our study on perturbed harmonic oscillators, having one and two hilltops and a plateau, demands that it requires a deeper understanding of such a behaviour. This study sheds light on how asymmetry in the potential impacts the properties of energy difference, eigenstates, and OTOCs. This study can also explain the exponential growth of the OTOC, even in the absence of any hilltop.

To see the role of asymmetry in potential, we consider unperturbed potentials having a hilltop, a plateau and two hilltops. In all the cases, a linear perturbation breaks the symmetry of potential landscape and eventually, for a certain perturbation, no hilltops appear. In all the cases, our OTOC calculations show that it grows exponentially for a range of eigenstates around the energy, which is equal to a classical turning point of the potential at which $ V^{\prime}(x) $ is a positive minimum. The difference in successive energy eigenvalues is minimum at this energy. The spreading of eigenfunctions and their probability distributions are the maximum around the point where  $ V^{\prime}(x) $ is a positive minimum. Our calculation of Loschmidt echo confirms the behaviour of OTOCs. These observations clearly indicate a strong correlation among exponential growth in OTOC, energy eigenvalue distribution, the structure of eigenstates, and asymmetry in the potential landscape.

The general observation is that the specific characteristics of OTOCs occur whenever the slope in the potential has a positive minimum. This naturally includes symmetric potential having a clear hilltop.

\appendix
\section{\label{Appendix-a}}
For a natural Hamiltonian of the form Eq.(\ref{eq3}), we have
\begin{equation}\label{A1}
	 \comm{\mathcal{\hat{H}}}{\hat{x}}=-2 i \hat{p}
\end{equation}.

 Applying $ \mel{m}{...}{n} $ to the both sides of the equation, we obtain,  
 \begin{equation}\label{A2}
\mel**{m}{ \comm{\mathcal{\hat{H}}}{\hat{x}}}{n}=-2 i\mel**{m}{ \hat{p}}{n}=-2ip_{mn}
 \end{equation}

\begin{eqnarray}\label{A3}
\Rightarrow -2ip_{mn}&=&\mel**{m}{ \comm{\mathcal{\hat{H}}}{\hat{x}}}{n}\\ \nonumber
&=&\mel**{m}{ (\mathcal{\hat{H}}\hat{x}-\hat{x}\mathcal{\hat{H}})}{n}\\ \nonumber
&=&	\mel**{m}{ \mathcal{\hat{H}}\hat{x}}{n}-\mel**{m}{ \hat{x}\mathcal{\hat{H}}}{n}\\ \nonumber
&=&	E_{m}\mel**{m}{\hat{x}}{n}-E_{n}\mel**{m}{ \hat{x}}{n},\qquad (\because \mathcal{\hat{H}}^{\dagger}=\mathcal{\hat{H}} \;\&\; \mathcal{\hat{H}}\ket{m}=E_m\ket{m})\\ \nonumber
&=&	(E_{m}-E_{n})\mel**{m}{ \hat{x}}{n}\\ \nonumber
\Rightarrow -2ip_{mn} &=&	E_{mn}x_{mn}\\ \nonumber
\Rightarrow p_{mn} &=& \dfrac{i}{2}	E_{mn}x_{mn}\\ \nonumber
\end{eqnarray}

\bibliography{Bibliography}

\end{document}